\documentclass[preprint]{aastex} 

\usepackage{graphicx}
\def\icarus{Icarus}

\def\degree{\ifmmode {^\circ}\else {$^\circ$}\fi}

\newcommand{\lae}{\lower 2pt \hbox{$\, \buildrel {\scriptstyle <}\over {\scriptstyle\sim}\,$}}
\newcommand{\gae}{\lower 2pt \hbox{$\, \buildrel {\scriptstyle >}\over {\scriptstyle\sim}\,$}}
\def\sgn{{\rm sgn}}

\def\eqnum[#1]{(\ref{#1})}

\def\mearth{\ifmmode {\rm M_{\oplus}}\else $\rm M_{\oplus}$\fi}
\def\mjup{\ifmmode {\rm M_{J}}\else $\rm M_{J}$\fi}
\def\msun{\ifmmode {\rm M_{\odot}}\else $\rm M_{\odot}$\fi}

\def\tsynodic{T_{\rm syn}}

\def\dadtfast{\frac{da_{\rm fast}}{dt}}

\def\dadtembedded{\frac{da_{\rm emb}}{dt}}
\def\dadtembedgap{\frac{da_{\rm gap}}{dt}}
\def\dadtvisc{\frac{da_{\rm visc}}{dt}}
\def\taufast{\tau_{\rm fast}}
\def\tauembedded{\tau_{\rm emb}}
\def\tauembedgap{\tau_{\rm gap}}
\def\tauvisc{\tau_{\rm visc}}

\def\dsp{\displaystyle}

\def\vkep{v_{_{\rm Kep}}}

\def\rhill{r_{_{\rm\!H}}}
\def\rfast{{{r}_{\rm fast}}}
\def\rgap{{{r}_{\rm gap}}}

\def\rphys{{r_{\rm phys}}}

\def\vice{v_{\rm p}}
\def\rice{r_{\rm p}}
\def\mice{m_{\rm p}}
\def\eice{e_{\rm p}}

\def\rhoice{\rho_{\rm p}}
\def\tclear{T_{\rm clear}}
\def\tfill{T_{\rm fill}}
\def\tdamp{T_{\rm damp}}

\def\floss{f_{\rm loss}}
\def\nurad{\nu_{\rm rad}}
\def\lamrad{\lambda_{\rm rad}}

\def\a130{\mbox{\small $1.3\!\times\!{10^5}$~km}}
\def\rkm{\mbox{\small 1~km}}
\def\rhoone{\mbox{\small 1~g/cm$^3$}}
\def\Sig40{\mbox{\small 40~g/cm$^2$}}
\def\noo85{\mbox{\small 85~cm$^2$/s}}
\def\h10{\mbox{\small 10~m}}
\def\rmeter{\mbox{\small 1~m}}

\def\codename{{\it Orchestra}}

\begin{document}

\title{Migration of small moons in Saturn's rings}

\author{Benjamin C. Bromley}
\affil{Department of Physics \& Astronomy, University of Utah, 
\\ 115 S 1400 E, Rm 201, Salt Lake City, UT 84112}
\email{bromley@physics.utah.edu}
\author{Scott J. Kenyon}
\affil{Smithsonian Astrophysical Observatory,
\\ 60 Garden St., Cambridge, MA 02138}
\email{skenyon@cfa.harvard.edu}

\begin{abstract}

The motions of small moons through Saturn's rings provide excellent
tests of radial migration models.  In theory, torque exchange between
these moons and ring particles leads to radial drift. We predict that
moons with Hill radii $\rhill \sim 2$--24~km should migrate through
the A~ring in 1000~yr.  In this size range, moons orbiting in an empty 
gap or in a full ring eventually migrate at the same rate.  Smaller 
moons or moonlets -- such as the propellers \citep[e.g.,][]{tis06} -- 
are trapped by diffusion of disk material into corotating orbits, 
creating inertial drag.  Larger moons -- such as Pan or Atlas -- do 
not migrate because of their own inertia.  Fast migration of 2--24~km 
moons should eliminate intermediate-size bodies from the A~ring and 
may be responsible for the observed large-radius cutoff of 
$\rhill \sim 1$--2~km in the size distribution of the A ring's 
propeller moonlets.  Although the presence of Daphnis ($\rhill \approx$ 5~km) 
inside the Keeler gap challenges this scenario, numerical simulations 
demonstrate that orbital resonances and stirring by distant, larger 
moons (e.g., Mimas) may be important factors. For Daphnis, stirring by 
distant moons seems the most promising mechanism to halt fast migration.  
Alternatively, Daphnis may be a recent addition to the ring that is 
settling into a low inclination orbit in $\sim 10^3$~yr prior to a phase 
of rapid migration. We provide predictions of observational constraints
required to discriminate among possible scenarios for Daphnis.

\end{abstract}

\subjectheadings{planetary systems -- solar system: formation}

\maketitle

\section{Introduction}


Migration is the steady radial drift of a massive planet or satellite
through an astrophysical disk \citep[e.g.,][and references therein]
{lin86,war97,art04,lev07,pap07,kir09,dang2010b,lub10}.  When a planet
lies embedded in a disk of gas or solid particles, it tries to clear
material from its orbit. Clearing by a massive planet creates wakes
inside and outside its orbit. These density perturbations exert a
torque on the planet, which produces an inwards or outwards migration
through the disk.  Analytic and numerical calculations show that, in
many cases, isolated planets migrate through the disk on a timescale
much shorter than the disk lifetime
\citep{gt79,lin79,gt80,ida00,tan02,pap07,kir09}.

This process provides a popular explanation for the architectures of 
many planetary systems. Migration plausibly accounts for the orbits 
of some ``hot Jupiters'' very close to their parent stars 
\citep[e.g.,][and references therein]{mas03,alex2009} and 
the compact orbital configurations of several multiplanet systems, 
including Kepler~11 which has five 
super-Earths orbiting within 0.3~AU of a Sun-like star \citep{lis11}. 
In our own solar system, migration appears to be a key dynamical 
mechanism for arranging the orbits of the outer planets 
\citep{mal93,hah99,lev07,mor07}.

Despite its broad applicability in exoplanetary systems, testing the
predictions of migration theory is challenging. In a young planetary
system, growing protoplanets gravitationally stir particles in the 
disk and limit the formation of coherent density wakes \citep{bk11b}. 
For a single planet, migration through a gaseous disk also depends 
on the dynamical and thermal state of the disk \citep[e.g.,][]{paa10,paa11}. 
Real disks contain several planets in a constantly evolving mix of gas 
and solid particles \citep[e.g.,][and references therein]{bk11a,youdin2012}.  
Current calculations only partially address this complexity
\citep[e.g.,][]{mas2001,mor07,dang2012}.  Thus, theory is hard-pressed 
to derive robust predictions applicable to observed systems.

Saturn's rings provide an interesting laboratory in which to test
migration theory. The A~ring is a geometrically thin disk of solid
particles with orbital periods shorter than a terrestrial day.  Aside
from countless cm- to m-sized particles \citep[e.g.,][]{cuz09}, the
rings contain many 0.1 to 1~km moonlets that could easily migrate
through the disk \citep{tis10}.  NASA's Cassini-Huygens missions have
also revealed a variety of density perturbations within the rings
\citep{col09}, identified a herd of ``propeller'' moonlets
\citep[named for the wakes of ring particles they
  create;][]{tis06,tis08}, and discovered the small moons Pan and
Daphnis \citep{sho91,por05}.  Together, these features allow robust
tests of our understanding of gravity and dynamics in a complicated
system of solid particles \citep[see also][and references therein]{cuzz2010,espo2010}.

Previous theoretical analyses of migration in Saturn's rings have
concentrated on the non-Keplerian, radial motion of propeller moonlets
within the A ring \citep{tis10}. These motions may result from
interactions between moonlets and stochastic density fluctuations in
the disk \citep{cri10,rei10,pan12b} or between moonlets and ring
particles on loosely bound ``frog'' orbits around the moonlet
\citep[][]{pan10,pan12a}.  These phenomena, or perhaps direct
collisions between moonlets and large ring particles, may {\em cause}
the observed radial motion in some moonlets. On the other hand,
large-scale ring structure, particularly radial density waves, may
provide torques that {\em suppress} radial motion \citep{tis12}. In
this picture, the observed non-Keplerian orbits come from recent kicks
received by moonlets, which then settle back to their original
Keplerian configuration.  Fortunately, each of these theoretical
scenarios makes a unique prediction for the evolution of moonlet
orbits. Upcoming Cassini observations will be able to distinguish
among them \citep{tis12}.

Here, we focus on other modes of migration for moonlets and small
moons embedded in Saturn's 17,000~km wide A ring\footnote{In the
  spirit of the definitions in \citet{tis13}, we use ``moonlet'' 
  to describe any of a numerous and perhaps unresolved population 
  with physical radii of $\lae$1~km and ``small moon'' to refer to 
  the larger objects Daphnis, Pan or Altas.}. Aside from stochastic
interactions, larger moonlets and small moons like Daphnis are
candidates for ``fast'' migration modes that satellites experience in
a dynamically cold, planetesimal disk \citep{ida00}. Because fast
migration is a critical component of migration theory in exoplanetary
systems, understanding fast modes in Saturn's rings improves our
ability to predict migration in a broad range of environments.

We begin with a theoretical overview of migration in a particle disk
(\S2) and then specialize the theoretical estimates to conditions
appropriate for Saturn's A~ring (\S3). In \S4, we confirm the analytic
estimates with numerical simulations, but demonstrate that stochastic
effects, resonances with more distant moons, and stirring by distant
moons can slow down or halt migration of small satellites through the
ring. In \S5, we examine the migration of Daphnis, which is located
within the Keeler gap.  If Daphnis only interacts with ring particles,
its migration should be fast and detectable with current
satellites. However, orbital resonances (e.g., Prometheus 32:31) and
stirring by moons out of the ring plane (e.g., Mimas and Tethys) may
have a strong impact on Daphnis' orbit and migration rate. Measuring
the evolution of Daphnis' orbit can constrain these possibilities. We
conclude in \S6 with a brief summary.

\section{Migration in a particle disk}
\label{sect:theory}

To test migration theory in Saturn's rings, we develop basic
predictions for an idealized disk of solid particles orbiting a much
more massive planet. For any satellite within the disk, the Hill
radius, $\rhill$, defines a spherical surface where the gravity of the
satellite roughly balances the gravity of the planet. This radius also
characterizes the annular width of the satellite's corotation zone, an
annulus where disk particles on nearly circular orbits oscillate back
and forth in the satellite's rotating reference frame (e.g., on
horseshoe orbits). Here we take the halfwidth of the corotation zone
to be $X_{co}\rhill$, with $X_{co} = 2$ \citep[e.g.,][]{bk11b}.  When
the corotation zone is full of small particles, migration is not
efficient.

Two physical processes set the density of smaller particles in the 
corotation zone.  The massive satellite gravitationally scatters smaller 
particles out of the zone. Viscous spreading drives the diffusion of 
smaller particles back into the zone. When scattering overcomes diffusion, 
the satellite opens up a physical gap in the disk. For convenience, we 
define $\rgap$ as the Hill radius of the smallest satellite capable of 
clearing out its corotation zone \citep[see also][]{lis81,raf01}.  When 
the corotation zone is nearly clear of smaller particles, migration can 
be very efficient. 

When a satellite has $\rhill > \rgap$, it undergoes fast migration. 
In this situation, the satellite can gravitationally pull corotating 
material {\em all the way past} its corotation zone, hauling itself 
radially inward or outward like a rope climber. In gaseous disks, 
this mode of migration is often called `type III' \citep{mas03} to 
distinguish it from the slow, type I migration of planets incapable
of forming a gap and the intermediate pace of type II migration of
more massive planets surrounded by a large gap\footnote{ In a gaseous 
  disk, uncertainties is the disk viscosity (and hence the radial and 
  vertical gradients in the pressure, temperature, and other physical
  variables) make migration calculations very challenging. Thus, it is
  uncertain whether type III migration is possible in a gaseous disk.
  In a particle disk, though, Type III migration is generally accepted. 
  Starting with \citet{mal93} and continuing with \citet{ida00}, 
  \citet{kir09}, and \citet{bk11b}, fast/rapid/runaway migration is 
  a standard phenomenon in particle disks.} \citep{war97,bry99}.

For more massive satellites with $\rhill \gg \rgap$, fast migration stalls.
As the mass of a satellite grows, its Hill radius -- and the width of its
corotation zone -- also grows. At some point, the satellite cannot scatter
particles out of the corotation zone fast enough: some disk material 
remains inside the corotation zone and interacts with the planet. This
interaction reduces the torque and slows migration. We define $\rfast$
as the Hill radius where fast migration begins to stall. Thus, satellites with
$\rgap < \rhill < \rfast$ undergo fast migration.

In physical units, the three critical Hill radii for fast migration are:
$$
\hspace{0.5in}
\begin{array}{ccllr}
\rhill & \equiv & \dsp a\, \left(\frac{m}{3M}\right)^{1/3}, 
& \mbox{[\small Hill radius]},
& \refstepcounter{equation}\hspace{0.5001in}(\theequation)
\label{eq:rhill}\\[10pt]
\rgap & \approx & \dsp 0.4 \ (\nurad a T)^{1/3} ,
& \mbox{[\small minimum $\rhill$ to form a gap]}, 
& \refstepcounter{equation}\hspace{0.5001in}(\theequation)
\label{eq:rgap}
\\[2pt]

\rfast & = & \dsp 1.7\,a\,\left(\frac{a^2\Sigma}{M}\right)^{\!{1}/{2}} 
& \mbox{[\small maximum $\rhill$ for fast migration]}. 
& \refstepcounter{equation}\hspace{0.5001in}(\theequation)
\label{eq:rfast}
\end{array}
$$ 
In these expressions, $m$ is the mass of an object orbiting with semimajor 
axis $a$ and period $T$ around a central star or planet with mass $M$. The 
disk has surface density $\Sigma$ and radial viscosity $\nurad$.

Deriving $\rgap$ and $\rfast$ relies on identifying important
timescales in the disk.  Rates of migration are set by the satellite's
Keplerian orbital period $T$.  Although an oblate planet and
gravitational perturbations by surrounding massive moons prevent a
satellite from executing standard Keplerian orbits, the Keplerian
orbital period is a useful fiducial for all other timescales.  Close
encounters between the satellite and particles occur every synodic
period $\tsynodic$, the timescale for particles on orbits with
separation $\Delta a$  to return to the same orbital phase.  
Massive satellites scatter particles away from their orbits, opening 
a gap of half-width $\Delta a$ within the disk clearing time, $\tclear$.  
For particle disks, the disk clearing time for a small gap is a few times
larger than $\tsynodic$ \citep[e.g.,][]{kir09}.  Viscous processes
between disk particles fill gaps on the viscous (or gap-filling) time
$\tfill$, which depends on the disk's radial viscosity $\nurad$
\citep[e.g.,][]{lis81,hour1984}.

Physical collisions damp the velocities of disk particles on the
damping timescale $\tdamp$.  Collisional damping is a random walk-type
process that depends on $\lamrad$, the radial distance between
collisions, and $\floss$, the fractional loss of kinetic energy per
particle in a collision (equivalent to the square of the coefficient 
of restitution).  The mean-free-path $\lambda$ is related to the optical 
depth $\tau$ \citep[e.g.,][]{cook1964,gt78a}. In the low optical depth 
limit, a ring of monodisperse particles with radius $\rice$ has 
$\tau = 3 \Sigma / 4 \rho_{\rm p} r_{\rm p}$, where $\rho_{\rm p}$ is 
the particle mass density \citep{gt78a,sal10b}.  A more general, 
realistic treatment of optical depth requires consideration of the 
particle size distribution and coherent structures that may form as a 
result of self-gravity \citep[e.g.,][]{sal03,tis13}.

For $\Delta a \ll a$, these timescales are:
$$
\hspace{0.5in}
\begin{array}{ccllr}
T & \equiv & 2\pi\!\left(\frac{a^3}{G M}\right)^{\!1/2}
& \mbox{[\small orbital period]},
& \refstepcounter{equation}\hspace{0.5001in}(\theequation)
\label{eq:torbit}\\[6pt]

\tsynodic & \approx & \frac{2 a T}{3 \Delta a} 
& \mbox{[\small synodic period for orbital separation $\Delta a$]},
& \refstepcounter{equation}\hspace{0.5001in}(\theequation)
\label{eq:tsynodic}\\[6pt]

\tclear & \approx & 4 \tsynodic 
& \mbox{[\small gap clearing time (half-width $\Delta a$)]},
& \refstepcounter{equation}\hspace{0.5001in}(\theequation)
\label{eq:tclear}\\[6pt]

\tfill & \approx & \frac{(2\Delta a)^2}{\nurad}
&  \mbox{[\small gap filling/viscous diffusion]},
& \refstepcounter{equation}\hspace{0.5001in}(\theequation)
\label{eq:tfill}\\[6pt]

\tdamp & \approx & \frac{\lamrad^2}{\floss\nurad} 
\approx \frac{4\rice \rhoice a}{3\floss \vkep \Sigma} 
\ \ \ \ 
& \mbox{[\small damping time]}.
& \refstepcounter{equation}\hspace{0.501in}(\theequation)
\label{eq:tdamp}
\end{array}
$$ 

To derive the physical scales for fast migration, we compare the
timescales for the relevant processes. When $\tclear = \tfill$, a
satellite clears its corotation zone as fast as viscous spreading
fills the zone back up.  This equality defines $\rgap$, the 
smallest Hill radius for a satellite capable of opening a gap in 
the disk.  Balancing the torque from the satellite on disk material
with the torque from the disk on the satellite is roughly equivalent 
to setting $\tclear = \tfill$ and yields similar expressions for 
$\rgap$ \citep[e.g.,][]{lis81,hour1984,ward1989,war97,raf01}. 

Aside from establishing which satellites can open gaps in the disk, 
$\rgap$ sets a lower limit on the Hill radii of satellites capable 
of undergoing fast migration \citep{ida00,mas03,kir09,bk11b}.  When 
a satellite pulls disk material near the edge of its corotation zone 
across its orbit, it recoils in the opposite direction.  If the net 
recoil is large enough, the satellite does not interact with scattered 
particles when they pass by the satellite one synodic period later. 
Satellites with $\rhill < \rgap$ cannot clear their corotation zones 
and do not recoil enough to overcome the inertia of material along 
their orbits.  Satellites with $\rhill \gg \rgap$ do not recoil 
enough to avoid a second encounter with scattered particles. 


To determine the range of Hill radii for satellites that undergo fast 
migration, we identify the maximum Hill radius where the time for 
a satellite to move a radial distance equal to the width of its 
corotation zone is less than the synodic period of particles at the 
edge of the corotation zone \citep[see also][]{ida00,mas03,kir09,bk11b}.  
To derive this Hill radius, which we define as $\rfast$, we set 
$\tsynodic = \eta \rfast [da_{\rm fast}/dt]^{-1}$ where 
$da_{\rm fast}/dt$ is the fast migration rate\footnote{\citet{bk11a}
derive this rate from the recoil due to a single scattering event and 
an encounter frequency integrated over the width of the corotation zone.}
\citep[e.g., eq.~12 below; see also][]{ida00}. To avoid repeat encounters
with material on horseshoe orbits within the corotation zone, we set
$\eta = 2$. Although choosing $\eta$ = 3.5 would eliminate all possible
repeat encounters, material trapped on horseshoe orbits provides most of 
the drag on a drifting moonlet. Thus, we choose $\eta$ = 2.


For disks of solid particles, the viscosity $\nurad$ depends on kinematics.
>From \citet[][and references therein]{gt78a}, the viscosity is a function
of the optical depth, the orbital period, and the radial velocity 
dispersion of particles $\vice$: 
\begin{equation}
\nurad \approx \frac{\vice^2 T}{4\pi}\frac{\tau}{1+\tau^2} ~ .
\label{eq: nu}
\end{equation} 
When $\tau$ is small, $\nurad \propto \tau$;
when $\tau$ is large, $\nurad \propto \tau^{-1}$.

In addition to the surface density and the viscosity, the vertical scale height 
of the disk $h$ also plays an important role in satellite migration
\citep[e.g.,][]{lin79,gt80,war97,ida00,pap07,bk11b}. For a disk of particles, 
the scale height depends on the vertical velocity dispersion $v_{\rm z}$:
\begin{equation}
h = v_{\rm z} \Omega^{-1} ~ ,
\label{eq: h-vert}
\end{equation}
where $\Omega$ is the local angular velocity of particles. If disk
particles follow a Rayleigh distribution, $v_{\rm z} \approx \vice /
2$ \citep[e.g.,][and references therein]{oht92}.  The local (3-D)
viscosity then scales approximately as $\nu \propto h^2$ as in the
$\alpha$ prescription for turbulent gaseous disks \citep{sha73,pri81}.
The true relationship between $v_z$ and the mean 3-D velocity dispersion 
is more complicated, especially in Saturn's A ring where self-gravity 
wakes can increase velocities in the plane of the ring \citep{tis07}.

With these definitions and physical expressions for critical parameters,
the theoretical radial drift rates for the three modes of migration are:
$$
\hspace*{11pt}
\begin{array}{rclclr}
\dsp\dadtembedded 
& \approx & 
\dsp -64 \frac{\pi a^2 \Sigma}{M}
  \,\frac{\rhill^3}{h^2 a} 
  \,\frac{a}{T} & 
\rhill \lae h, \rgap & 
\mbox{[\small embedded, no gap; Type I]} 
& \refstepcounter{equation}\hspace{8.555pt}(\theequation)
\label{eq:dadtembedded}
\\[12pt]
\dsp \dadtfast & 
\approx & 
\dsp \pm \frac{5.3 \pi a^2 \Sigma}{M}
\frac{a}{T} 
& 
h,\rgap \lae \rhill \lae \rfast
& \mbox{\small[fast; Type III]}
& \refstepcounter{equation}(\theequation)
\label{eq:dadtfast}
\\[12pt]
\dsp
\dadtembedgap & \approx & 
\dsp
- 16 \frac{\pi a^2\Sigma}{M}\frac{\rhill}{a} 
\,\frac{a}{T} 
\rightarrow \ -\frac{3\nurad}{2a}
& 
\rhill \gae \rgap & 
\mbox{\small [embedded$+$gap; Type II]}.
& \refstepcounter{equation}(\theequation)\label{eq:dadtembgap}
\end{array}
$$ 
When the Hill radius is smaller than the disk scale height and the gap
radius, the satellite has a very weak gravity, cannot open up a gap,
and produces very weak density perturbations in the disk.  Only very
slow, embedded migration is possible.  Its rate
(eq.~[\ref{eq:dadtembedded}]) follows from weak scattering theory
\citep[e.g.,][]{lin79} and is analogous to Type~I migration in gaseous
disks \citep{war97}. The scale height is important in setting this
rate because it is a measure of the distance at which the ring
particles transition from random motion to coherent, shear flow from
the satellite's perspective \citep[e.g.,][]{tan02}. Fast migration
(eq.~[\ref{eq:dadtfast}]) is the ideal case of a satellite spiraling
through a dynamically cold disk \citep{mas03}. The rate is independent
of satellite mass; the greater gravitational reach of a more massive
satellite compensates for its higher inertia. The migration rate for
$\rhill>\rfast$ is the Type II analog for a planet which opens a gap 
in a particle disk (eq.~[\ref{eq:dadtembgap}]). In the low viscosity 
limit, differential torques arise from material at the edges of the 
gap, which lies at or beyond the corotation zone \citep{bk11b}. In 
the high viscosity limit, the migration rate is set by the viscous 
timescale \citep[see, e.g.,][]{war97,pap07}.

The transition from type I migration with no gap to fast and type II
migration with a gap is complex \citep[e.g.,][]{hour1984,ward1989}. 
When a small satellite with $\rhill < \rgap$ accretes disk material 
and grows to larger radii, it drifts through the disk more rapidly 
(eq. [\ref{eq:dadtembedded}]). If this satellite drifts across its
corotation zone more rapidly than it can clear a gap, it drifts through
the disk with a partially cleared corotation zone. Because the drift
time scales more weakly with mass ($\propto m^{-1}$) than the time to
clear a gap ($\propto m^{-2}$), there is a range of satellite masses 
where a satellite could clear a gap if it could stop drifting 
\citep{hour1984,ward1989}. This range of masses roughly corresponds to
the range where fast migration operates.

The transition from type I to type II migration also depends on the
scale height and eccentricity $e$ of disk particles.  When disk particles 
have large $e$, the drift rate slows by a factor $\sim(\rhill/\eice a)^3$ 
relative to the fast rate (see eq. [\ref{eq:dadtfast}]; Bromley \& Kenyon 2011b).  
When $h < \rhill < \rgap$, simulations suggest that nonlinearities and other
competing effects -- such as the inertial drag of material stuck in
the corotation zone -- substantially reduce the migration in this
``transgap'' range of Hill radii \citep{dan03}. This conclusion is
strengthened by a recent analysis of moonlets in the A ring \citep{tis12}, 
which shows how radial density structures in the ring can produce torques 
that damp radial drift.

For larger satellites with $\rhill \sim \rfast$, simulations \citep{bk11b} 
also suggest that fast migration falls off at least as rapidly as 
$(\rfast/\rhill)^3$ \citep[see also][]{sye95}. As 
$\rhill$ increases 
($\rhill > \rgap, \rfast$), the satellite can maintain a gap
in the disk and can experience migration with a fully-formed gap
(eq.~[\ref{eq:dadtembgap}]).  In the inviscid limit of this
type II migration \citep{war97,cri2006}, the rate is given by weak 
scattering theory. In a sufficiently viscous disk, material piles 
up at the gap edges. Then, the timescale for radial drift is the 
viscous timescale \citep{bry99}.

\begin{figure}[htb]
\centerline{\includegraphics[width=5.5in]{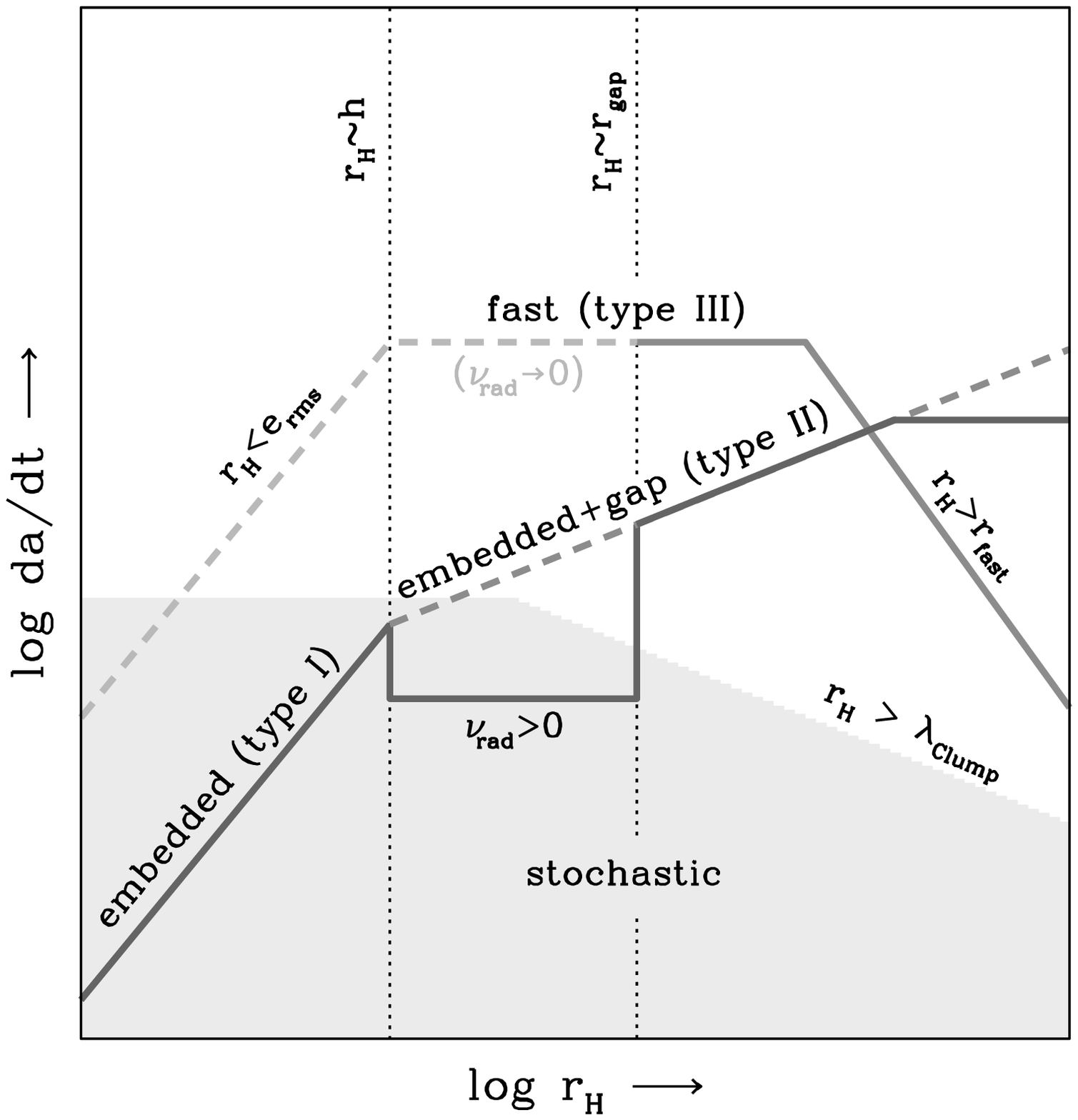}}
\caption{
A qualitative comparison of predicted migration rates as a function of Hill
radius for satellites in a particle disk. The solid curves show rates
in viscous disks; the dashed curves apply in the limit of zero
viscosity where $\rgap$ --- the Hill radius of a satellite that can
clear a gap --- is set by the disk scale height, $h$.  A satellite
drifts radially according to the larger of the fast and embedded
migration rates for its Hill radius.  The dip in the embedded
migration rate for $\nurad > 0$ illustrates that migration may be
suppressed in a viscous disk if $\rhill$ is in the ``transgap''
regime, where the satellite is larger than the disk scale height but
not large enough to open a gap. The curve representing fast migration
shows an upper plateau with a migration rate independent of $\rhill$,
a steep decline above the limiting Hill radius $\rfast$ (see text),
and a similarly sharp attenuation in $da/dt$ below $\rhill = h$,
where the typical disk particle eccentricity $e_{\rm rms}$ becomes important.
The shaded region is suggestive of radial drifts from stochastic
migration within some fixed time period. If the source of the
stochasticity is clumping of disk particles on scales of $\lambda_{\rm
  clump}$, then the effective rate falls off as $1/\rhill$. 
Otherwise the effective rate is constant.
\label{fig:migratesketch}}
\end{figure}

The sketch in Figure~\ref{fig:migratesketch} summarizes the various
migration modes. In addition to showing the relative magnitude of the
rates in equations (\ref{eq:dadtembedded}--\ref{eq:dadtembgap}) as a
function of $\rhill$, the sketch illustrates the impact of viscosity
in the transgap regime and the high- and low-mass fall off of the fast
migration mode. Other motions affecting a satellite's radial drift
include stochastic migration \citep[see][]{rei10,cri10,pan12b} and the
periodic orbit perturbations caused by material on corotating ``frog''
orbits centered on the satellite \citep{pan10}.  On yearlong
timescales, these drifts (random walks with effective $da/dt$ values
that lie in the shaded region in the figure)
can be comparable to linear-in-time radial
migration in Saturn's rings. 

\section{Migration in the A ring}
\label{sect:Aring}

Saturn's A ring is a dynamically cold, thin disk of icy, meter-sized
particles, lying in an annulus around Saturn that extends from a radius of
117,000~km to the Roche Division at 137,000~km.  In our calculations,
we adopt an orbital radius of $a = 130,000$~km as representative.
For other physical parameters, observations \citep[e.g.,][]{tis07,col09} 
suggest:
\begin{eqnarray}
\label{eq:Sigmafid}
\Sigma &=& 40 \ \mbox{g/cm$^2$} \\ 
\label{eq:hfid}
h &=& 10 \ \mbox{m} \\
\label{eq:nuradfid}
\nurad &=& 85 \ \mbox{cm$^2$/s}.
\end{eqnarray}
Adopting these values allows us to specialize the results of 
\S\ref{sect:theory} to Saturn's A ring.

In the A ring, the radii that delineate different modes of migration
in equations~(\ref{eq:rhill}--\ref{eq:rfast}) become
\begin{eqnarray}
\label{eq:rhillA}
\rhill &=& 1.7 \  
\left[\frac{\rho}{\rhoone}\right]^{\!{1}/{3}}
           \left[\frac{a}{\a130}\right] 
           \rphys,
\\
\label{eq:rgapA}
\rgap & \approx &  
2\, \left[\frac{\nurad}{\noo85}\right]^{\!{1}/{3}}
    \left[\frac{a}{\a130}\right]^{\!{5}/{6}}
\ \mbox{km}, 
\\
\label{eq:rfastA}
\rfast & \approx & 24 \,
    \left[\frac{a}{\a130}\right]^{2}
                \left[\frac{\Sigma}{\Sig40}\right]^{\!{1}/{2}}
     \,\mbox{km}. 
\end{eqnarray}
With mean densities of $\sim 0.5$--1~g/cm$^3$, the Hill
radius and the physical radius of a moonlet or small moon in the
A~ring are similar.  From equation~(\ref{eq:rgapA}), satellites with
Hill radii below a few~km cannot open a full gap in the A ring.  These
bodies can only migrate in the slow, embedded mode.  Satellites
with $\rhill$ $\gtrsim$ 20~km are too massive for fast migration.
Between a few km and 20 km, small moons can migrate in fast mode.

The timescales in equations~(\ref{eq:torbit}--\ref{eq:tdamp})
specialize to
\begin{eqnarray}
T & \approx & 0.0015 \, \left[\frac{a}{\a130}\right]^{{3/2}} \ {\rm yr},
\\
\label{eq:tsynodicA}
\tsynodic & \approx & 130 \, 
        \left[\frac{\rkm}{\Delta a}\right]
        \left[\frac{a}{\a130}\right]^{{5/2}} \ {\rm yr},
\\
\label{eq:tclearA}
\tclear & \approx & 520 \, 
        \left[\frac{\rkm}{\Delta a}\right]
        \left[\frac{a}{\a130}\right]^{{5/2}} \ {\rm yr},
\\
\label{eq:tfillA}
\tfill  & \approx &  3.7 \, \left[\frac{\Delta a}{\rkm}\right]^2
           \left[\frac{\noo85}{\nurad}\right]   \ {\rm yr},
\\
\label{eq:tdampA}
\tdamp  & \approx & 0.008 \, 
           \left[\frac{0.1}{\floss}\right]
           \left[\frac{\rho}{\rhoone}\right]
           \left[\frac{\Sig40}{\Sigma}\right]
           \bigg[\frac{\rice}{\rmeter}\bigg] 
           \left[\frac{a}{\a130}\right]^{{3/2}} \ {\rm yr} ~ .
\end{eqnarray}
In the expression for the damping time, we adopt $\floss = 0.1$ 
(see \S\ref{subsect:distmoons}).

Finally, the migration rates in the A ring, corresponding to
equations~(\ref{eq:dadtembedded}--\ref{eq:dadtembgap}), are:
$$
\begin{array}{rclcr}
 \dsp\dadtembedded 
 & \!\lae\! & 
\dsp
-10^{-4} \ 
   \bigg[\frac{\h10}{h}\bigg]^{\!2}
   \!\left[\frac{a}{\a130}\right]^{\!1/2} 
   \!\left[\frac{\Sigma}{\Sig40}\right]
\mbox{km/yr} 
&
 \rhill \lae {2}~\mbox{km} 
& \refstepcounter{equation}\hspace{8.555pt}(\theequation)
\label{eq:dadtembeddedA}
\\[12pt]
\dsp \dadtfast & 
\!\approx\! & 
\dsp \pm 17 \ 
   \left[\frac{a}{\a130}\right]^{\!{3}/{2}} 
   \left[\frac{\Sigma}{\Sig40}\right]
   \ \mbox{km/yr}
& 
2\ \mbox{km}\lae \rhill \lae 24\ \mbox{km}
& \refstepcounter{equation}(\theequation)
\label{eq:dadtfastA}
\\[12pt]
\dsp \dadtembedgap  & \!\approx\! &  
\dsp -0.009 \ 
   \bigg[\frac{\rhill}{\mbox{\small 24 km}}\bigg]\!
   \left[\frac{a}{\a130}\right]^{\!1/2}\!
   \left[\frac{\Sigma}{\Sig40}\right]
\,\mbox{km/yr}
&
\rhill \gae 24\ \mbox{km}\,.
& 
\refstepcounter{equation}(\theequation)\label{eq:dadtembgapA}
\end{array}
$$
Equation (\ref{eq:dadtembeddedA}) gives the maximum embedded rate
for $\rhill \sim h$, with the expectation that embedded
satellites with larger radii cannot migrate faster as a result of 
viscous effects \citep[e.g.,][]{dan03}.
Equation (\ref{eq:dadtembgapA}) expresses the migration rate in the
limit of small viscosity. In a viscous disk,
$$
\begin{array}{rclcr}
\dsp \dadtvisc  & \!\approx\! &  
\dsp -3 \times 10^{-6} \ 
   \left[\frac{\a130}{a}\right]\!
   \left[\frac{\nurad}{\noo85}\right]
\,\mbox{km/yr}
&
\rhill \gae 24\ \mbox{km}, \nurad \gg 0 ~ .
& 
\refstepcounter{equation}(\theequation)\label{eq:dadtvisc}
\end{array}
$$
These equations provide an idealized description of migration
in the A~ring.  At small $\rhill$, we take the embedded migration rate
for $\rhill \sim 3 h$ as a maximum value, since larger Hill radii are
in the transgap regime where migration is likely suppressed.  The 
large-$\rhill$, viscosity-dominated, Type~II migration rate is formally 
below 1~cm/yr.

\begin{figure}[htb]
\centerline{\includegraphics[width=5.5in]{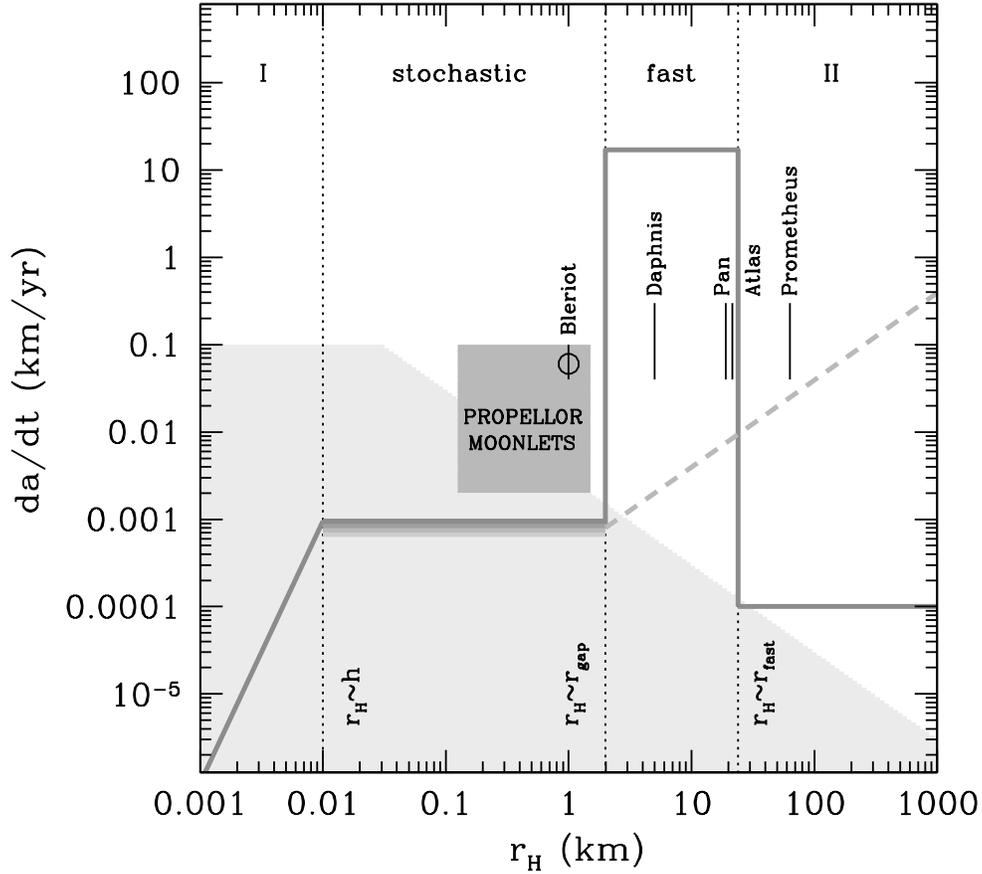}}
\caption{A schematic of predicted migration rates as a function of Hill
radius for satellites in Saturn's A ring. The solid curve shows
the upper limit of the predicted rate for a wide range
of satellite masses. The dashed line is the low-viscosity limit
of embedded migration. For $\rhill \gtrsim \rfast$, the solid 
line below the dashed line reflects the viscosity-dominated 
Type II mode. The shaded region represents stochastic migration,
bounded by an approximate average yearly radial drift assuming that the
ring contains clumps of 30~m (these rates are drawn from a broad range
reported by \citet{cri10} and \citet{rei10}).  
The data points and shaded region for the propeller moonlets indicate 
Hill radii only; with the exception of the moonlet Bl\'{e}riot, 
migration has not been detected.
\label{fig:migrateAring}}
\end{figure}

To put these results in perspective, we estimate the timescales for migration
through a distance of 10,000~km, approximately half the annular width of 
the A ring:
$$
\begin{array}{rclcr}
 \dsp\tauembedded 
 & \!\gae\! & 
\dsp
 100 \
   \bigg[\frac{\rkm}{\rhill}\bigg]^{\!3}
   \!\bigg[\frac{h}{\h10}\bigg]^{\!2}
   \!\left[\frac{\a130}{a}\right]^{\!1/2} 
   \!\left[\frac{\Sig40}{\Sigma}\right]
\ \mbox{Myr} 
&
 \rhill \lae {2}~\mbox{km} 
& \refstepcounter{equation}\hspace{8.555pt}(\theequation)
\label{dadtauembeddedA}
\\[12pt]
\dsp \taufast & 
\!\approx\! & 
\dsp 0.0006  \ 
   \left[\frac{\a130}{a}\right]^{\!{3}/{2}} 
   \left[\frac{\Sig40}{\Sigma}\right]
   \ \mbox{Myr}
& 
2\ \mbox{km}\lae \rhill \lae 24\ \mbox{km}
& \refstepcounter{equation}(\theequation)
\label{eq:taufastA}
\\[12pt]
\dsp \tauembedgap  & \!\approx\! &  
\dsp 1.1 \ 
   \bigg[\frac{\mbox{\small 24 km}}{\rhill}\bigg]\!
   \left[\frac{\a130}{a}\right]^{\!1/2}\!
   \left[\frac{\Sig40}{\Sigma}\right]
\,\mbox{Myr}
&
\rhill \gae 24\ \mbox{km}\,.
& 
\refstepcounter{equation}(\theequation)\label{eq:tauembgapA}
\\[12pt]
\dsp \tauvisc  & \!\approx\! &  
\dsp 370 \
           \left[\frac{\noo85}{\nurad}\right] 
\,\mbox{Myr}
&
\mbox{(viscous lifetime)}\,.
& 
\refstepcounter{equation}(\theequation)\label{eq:tauviscA}
\end{array}
$$
Figure~\ref{fig:migrateAring} summarizes results for the migration rate 
in the A ring as a function of Hill radius.  Satellites in or near the 
A ring are included in the diagram. 

These timescales lead to several clear conclusions regarding migration 
through the A ring. Objects with Hill radii less than roughly 0.25~km 
migrate on a timescale comparable to the age of the solar system
\citep[see also][and references therein]{gt82}. This radius is roughly 
the maximum Hill radius of propeller moonlets in the A ring.  Because 
the migration time scales as $\rhill^3$, fast migration may be responsible 
for the lack of propeller moonlets with $\rhill \gtrsim$~1~km.  Two moons, 
Pan and Atlas, have Hill radii very close to the fast migration cut-off 
$\rfast$.  Although the predicted lifetimes for these moons are short 
compared to the age of the solar system, they are very long compared to 
the fast migration lifetime. 

The small moon Daphnis clearly provides a more extreme challenge to
migration theory. Its radius lies right in the middle of the expected
range of radii for fast migration. With an expected migration time of
less than 1000 yr, either (i) Daphnis is a recent addition to the rings,
(ii) the migration theory is incorrect, or (iii) additional factors 
contribute to its apparent lack of migration. In the following sections, 
we address these possibilities and their impact on the rates shown in 
Figure~\ref{fig:migrateAring}.

%
%

\section{Migration of small moons: N-body experiments}
\label{sect:numerics}

To explore the migration of small moons through Saturn's A ring in more
detail, we examine a set of $N$-body calculations. To perform these
calculations, we use \codename, a hybrid $N$-body-coagulation code 
which includes additional modules for the radial diffusion of gas and
solid particles through a disk surrounding a star or a planet. In
\citet{bk06}, \citet{kb04,kb08,kb09}, and \citet{bk11a}, we describe 
each component of \codename, including comprehensive results with 
standard test problems. Here, we provide a short overview, with 
a focus on the $N$-body component, which we use exclusively in the 
calculations for this paper. 

\subsection{N-body simulation method}
\label{subsect:nbodymethod} 

The $N$-body algorithm distinguishes among three types of particles:
the massive $N$-bodies, the ``swarm'' (superparticles), and the tracers. 
The massive $N$-bodies gravitationally interact with all particles in a
simulation. The swarm particles interact with and can influence the
motions of the massive $N$-bodies but do not interact with each other
or with the tracers.  The tracers respond to the gravity of the massive 
$N$-bodies. Here, we use massive $N$-bodies to track the motions of 
Saturn, its major moons, and the smaller moons.  Unless noted otherwise, 
we represent the gravity of the ring using swarm particles.  Thus, the 
ring is not self-gravitating.

\codename\ uses a 6$^{\rm th}$-order accurate symplectic integrator
for all massive bodies in a computational domain that encompasses the
full 3-D extent of all orbits. In the case of the relatively numerous
tracer or swarm particles, a software switch allows for 4$^{\rm
  th}$-order accurate integration over a timestep that is typically
1/300$^{\rm th}$ of the orbital period. During this step, the orbits of
massive particles are interpolated in a way that allows them to respond 
to the gravity from swarm particles. The code handles close encounters 
with a massive body by solving Kepler's equations.  These features allow 
us to track up to $10^8$ ring particles for thousands of orbits, or 
$10^5$ ring particles for millions of orbits.

\codename\ can also accommodate rates of change in orbital elements, 
given by $da/dt$ (semimajor axis), $de/dt$ (eccentricity), and $di/dt$ 
(inclination), from physical interactions among the swarm particles.  
Here, we use this feature to include the evolution of solid ring 
particles in a viscous disk.  To calculate changes in these elements 
over a time step $\Delta t$, we use
\begin{eqnarray}
\Delta a &=& \sgn(u-0.5) \sqrt{\nurad \Delta t} \\
\Delta e &=& -e \frac{\Delta t}{\tdamp} \\
\Delta i &=& -i \frac{\Delta t}{\tdamp}
\label{eq:damping}
\end{eqnarray}
where $u$ is a standard uniform (pseudo)random variate used to
implement a random walk to mimic the radial diffusion of the swarm
particles.  For individual isolated particles scattered within roughly 
a Hill radius of a small moon, the code has a switch to turn off this 
feature.

In the calculations for this paper, we chose a damping time of $0.01$~yr, 
comparable to the predicted timescale in equation~(\ref{eq:tdampA}).  The
diffusion of small ring particles is an important aspect of the
interactions between massive $N$-bodies and the swarm.  As long as it
is within an order of magnitude of our choice, the precise value of
$\tdamp$ does not make any noticeable difference in the outcomes of
our simulations,

To test this aspect of our $N$-body code, we considered the evolution 
of a narrow ring of small particles evolving solely by diffusion and 
collisional damping. Starting with initial conditions identical to 
\citet{sal10b}, our simulations yield a gradual expansion of the narrow 
ring into a broad disk. For at least 100~Myr, the time evolution of the 
surface density distribution in our simualtions closely follows 
\citet{sal10b}, who solve the 1D radial diffusion equation for a disk of 
small particles. Comparisons with our own solutions to the radial diffusion 
equation further demonstrate that the $N$-body code treats particle 
diffusion and collisional damping accurately. 

Finally, we developed and tested algorithms to model Saturn's 
oblateness, which can significantly affect orbital phases and 
inferred orbital elements for all satellites and ring particles. 
When we include Saturn's oblateness in the calculations, we set the 
gravitational acceleration at a distance $\vec{x}$ near the ring to 
\begin{equation}
\label{eq:oblate}
\frac{d^2\vec{x}}{dt^2} = - \frac{G M}{x^3} 
    \left[1 + {3} J_2 P_2(\sin \phi)\frac{R_S^2}{x^2}
    \right] \vec{x},
\end{equation}
where $\phi$ is the latitude relative to the ring, $P_2$ the
second-order Legendre polynomial, and the gravitational constant $J_2
= 0.01629$ for an adopted radius of Saturn $R_S = 60,330$~km
\citep{jac06}.

For this paper, we consider two sets of simulations.  When examining a
simple physical process -- such as a small moon migrating through a pristine 
ring -- we treat the dynamical interactions between a small moon and ring 
particles using swarm particles and the damping/viscous interactions 
among ring particles (eq. [\ref{eq:damping}]), but we do not generally
include the oblateness of Saturn or interactions with distant moons 
(e.g., Mimas, Rhea, Dione, etc).  Eliminating oblateness and the distant 
moons allows us to investigate a larger range of input parameters per 
unit cpu time. Once we establish a typical behavior, we verify that 
including oblateness and interactions with distant moons do not change
this behavior.  In more complex simulations (e.g., for Daphnis in \S5),
we include oblateness and the interactions with distant moons and 
note the included moons in the text.

\subsection{Comparison with other techniques}

The $N$-body algorithm in \codename\ differs from `patch' simulations
of Saturn's rings, where the computational domain is a comoving
rectilinear patch centered on a circular Keplerian orbit in the ring
\citep[e.g.,][]{lew00,por08,salo10,lew11,per11}. Patch calculations
include $10^5$ to $10^6$ ring particles with typical masses of $5-10
\times 10^{-5}$ relative to the satellite of interest.  For patches
with sizes of roughly a square kilometer, these calculations can track
the interactions and long-term evolution of ring particles with
realistic sizes, $\sim$ 1--10~m.  To understand the origin of local
structures within the rings, these calculations are ideal. However,
periodic boundary conditions at the 'ends' of a patch limit the
ability of these calculations to include perturbations from material
outside the patch. Thus, it is challenging for these simulations to
track the response of a moonlet or small moon to the global evolution
of ring particles.

In \codename, the $N$-body code follows much larger particles, $\sim$
10--100~m, over complete orbits around the central mass.  As long as
the satellite mass exceeds the ring particle mass by a factor of a few
hundred, calculations suggest that these superparticles can accurately
represent smaller ring particles and their effect on an embedded
satellite \citep[Fig.~\ref{fig:ringxx} below; see also][]{kir09}. This
approach also enables straightforward inclusion of resonances and
other perturbations from distant moons. In our simulations, the number 
of superparticles is too small to identify small-scale or short-term
structures within the ring particles. Thus, these calculations
accurately follow the evolution of a small moon at the expense of a
large graininess in the structure of the ring. Because our focus is on
the long-term evolution of a satellite in response to the entire ring
and distant moons, this compromise is reasonable.

Although most previous $N$-body work has focused exclusively on ring
particle dynamics, several studies consider radial drift of satellites
within or outside the ring \citep{lew09,cha11,tis12}.  Despite their
success in identifying sources of radial drift, none of these
simulations can track fast migration.  For example, \citet{cha11}
elegantly track Type II-like migration by summing the torque on a
moonlet at Lindblad resonances.  However, the relevant part of the
ring for fast migration is closer to the corotation zone than the
Lindblad resonance \citep[see][for the analogous case of Type III
  migration in a gaseous disk]{mas03}.  Thus, this approach will never
yield fast migration.

\citet{lew09} provide an excellent demonstration of chaotic drift of a
moonlet by collisions and interactions with massive particles and
self-gravitating wakes in the rings. However, their patch-like
computational domain spans only a small fraction of a moonlet's orbit
around the planet. To produce fast migration in this type of
calculation, we speculate that it is necessary to account for the
phase and timing of a ring particle's orbit when it reenters the
patch.  In all of their calculations, the moonlets of interest to
\citet{lew09} are too small to open the annular gap in the ring
necessary for fast migration (e.g., $\rhill < \rgap$).  Thus, fast
migration is not expected in any of their simulations.

\subsection{Onset of fast migration}


Sustained fast migration of a small moon requires one simple condition. 
The moon must be able to draw/scatter ring material from one edge of its 
corotation zone to the other and then drift a sufficiently large radial 
distance in the opposite direction to avoid encountering the scattered 
particles again.  Moons that are too large cannot drift far enough before 
re-encountering scattered particles.  Very small moons cannot clear enough 
of an annular gap and are inertially pinned by other corotating material 
at smaller orbital separation.  Thus, moons which are too small or too 
large migrate very slowly.




Despite this clear picture for {\it sustaining} fast migration, {\it initiating}
fast radial drift is more subtle. The onset of fast migration requires 
(i) a satellite with a fairly empty corotation zone and (ii) enough 
disk material close by, within $\sim 3 - 4 \rhill$, where dynamical
interactions can draw material into the satellite's corotation 
zone\footnote{In addition to particles on horseshoe orbits, material on 
initially circular orbits with $\Delta a$ as large as $2\sqrt{3}\rhill$
can be deflected onto crossing orbits with the satellite \citep{gla93}.}. 
Among plausible starting points, we consider two extreme initial 
conditions, a satellite lying in a completely empty gap in the disk
and a satellite embedded in a continuous disk.

Although a satellite within a continuous disk has enough material 
nearby to begin fast migration, it must first clear its corotation 
zone. For satellites with $\rhill \gtrsim \rgap$, the timescale for 
scattering particles out of the corotation zone is smaller than the
timescale for viscous diffusion to bring particles back into the
corotation zone. Thus, the satellite begins to clear its corotation 
zone \citep[see also][]{lis81,hour1984,raf01}. Whether or not the 
satellite drifts slowly in the type I mode, any imbalance in the net 
amount of scattering into the inner disk relative to the outer disk 
creates a net drift sufficient to begin fast migration. As long as 
the physical properties of the disk and the satellite maintain the 
conditions $\tclear \gtrsim \tfill$ and $\rhill \lesssim \rfast$, 
the satellite continues fast migration.

For a satellite within a large gap, the onset of fast migration is
more delicate. Despite its empty corotation zone, the satellite must
reach across the gap and draw material from the gap edges into the
region, $\Delta a \lesssim 2\sqrt{3}\rhill$ $\sim 3.5 \rhill$, where 
orbit crossings can begin. For a satellite with $\rhill \gg \rfast$, 
the half-width of the gap is much larger than $\sim$ 3.5 $\rhill$. Disk 
material is too far away; the satellite can only migrate in the type~II 
mode. For satellites with $\rgap \lesssim \rhill \lesssim \rfast$, the 
edges of the gap are within $\sim$ 3.5 $\rhill$. Material can then be 
drawn into crossing orbits with the satellite.

Although satellites with $\rgap \lesssim \rhill \lesssim \rfast$ have
fairly small gaps, it is possible to construct a disk where orbit
crossings never occur.  We first consider a disk where the surface
density at the inner edge of the gap is comparable to the surface 
density at the outer edge. When (i) a disk has a vertical scale height 
$h \equiv$ 0, (ii) the satellite lies on a perfectly circular orbit 
centered in a gap with half-width $\Delta a \gtrsim 4 \rhill$, and 
(iii) there are no external torques, the torque balance between the 
satellite and the gap edges is perfect \citep[e.g.,][]{hour1984}.  
The satellite cannot draw material into its corotation zone and can 
only migrate slowly in the type II mode.  

In a real disk, the finite scale height breaks the symmetry, allowing the 
satellite to draw material from the gap edges \citep{hour1984}.  Once the 
satellite begins to draw material from the edges of the gap across its 
corotation zone, additional jostling of the gap edges or the satellite's 
orbit -- for example, by density fluctuations in the disk or by gravity 
fluctuations due to distant moons -- places the satellite closer to 
material at one edge of the gap than the other edge. The satellite draws 
more material into its corotation zone. Fast migration commences.

To try to maintain the satellite on a stable orbit, we can specify the 
surface density on the inner/outer edges of the gap to balance the slightly
different angular momenta of particles on either edge of the gap. We can
also specify that the number of orbit crossings from material in the inner 
disk precisely balance the number from the outer disk. Every ensemble of
scattering events then leads to no radial motion. Neglecting internal ring 
dynamics and other external processes, the satellite can maintain a wide, 
clear gap and remain on its initial orbit indefinitely. 

In a real ring, however, this balance is metastable. Radial diffusion tries 
to fill the gap. Perturbations from scattering by self-gravitating wakes, 
direct collisions, or asymmetries in the Type I-like torques 
(eq.~[\ref{eq:dadtembedded}]) from more distant ring material pushes 
ring particles into the gap \citep[e.g.,][]{tis12}. Jostling by distant
moons also tends to fill the gap. If these processes fill both sides of
the gap identically, it is possible to maintain the balance. If these
perturbations leave a few extra particles on one edge of the gap, the
additional crossing events will pull the moon in that direction. Once
the moon is drawn to either gap edge, it will pull even more particles
from that edge, enhancing its radial drift. Fast migration again commences.

To explore these two possibilities in more detail, we examine a suite
of numerical simulations for satellites in disks with physical 
characteristics similar to Saturn's A ring. We begin with a discussion
of fast migration for a satellite in a completely empty gap. We then
consider a satellite in a continuous disk.




\subsubsection{Fast migration in smooth rings}

To illustrate metastability of a satellite in a pristine gap, we set up 
an initially stable configuration with balanced torques from either edge 
of the gap. The simulations consider two plausible initial conditions for 
particles spaced uniformly in radius through the ring: particles spaced 
(i) randomly or (ii) uniformly in the azimuthal direction. The 
calculations also consider two types of ring particles: (i) massive swarm 
particles, where we calculate the radial drift from individual $N$-body
interactions directly or (ii) massless tracer particles, where we derive
the drift from the change in the total angular momentum of the tracers.
This approach allows us to test the sensitivity of the migration rate 
to the initial orbits of ring particles and to the method for calculating
torques between the moon and ring particles.

For a small moon with $\rhill = 7$~km and a gap with half-width = 2 $\rhill$, 
fast migration initiates in every simulation in this suite of tests 
(Fig. \ref{fig:ringxx}). Moons interacting with tracer particles tend to 
begin migrating earlier than moons surrounded by massive swarm particles. 
Moons within a random distribution of ring particles appear to start migrating 
earlier than moons in a uniform distribution. In all cases, the onset of fast 
migration requires a short time, smaller than the synodic period of material 
at the gap edge. When the mass ratio of the moon to a ring particle is large, 
$>10^3$, the onset is slow and regular. When the mass ratio is $\lesssim 10^2$, 
migration is more stochastic and start more quickly due to individual 
scattering events that push the moon toward the edge of the gap.  Still, both 
sets of initial conditions, both approaches to calculate the radial drift, and 
all mass ratios yield roughly the same migration rate.



To explore the onset of fast migration in more detail, we conducted
additional tests with larger gaps and different diffusion
timescales. The range of gap sizes, with half-widths $\Delta a$ from
$2\rhill$ to $6 \rhill$, includes two nominal half-widths that are
important from gap-formation theory (see the Appendix): the first is
$\Delta a \approx 3 \rhill$, derived by balancing torques between the
satellite and the disk \citep{lis81}, while the second is $\Delta a
\approx 2.5 \rhill$, derived by balancing scattering with viscous
diffusion.  In all these tests, ring particles slowly diffuse into the
gap.  The width of the gap around the moon gradually decreases.  With
the nominal diffusion time of 10~yr for a small, gap-clearing moon,
larger gaps fill in more slowy, and imply longer times before
migration can start.  When the gap width closes to roughly twice the
corotation half-width, $\Delta a \approx 4 \rhill$, the moon starts to
chaotically scatter ring particles. Fast migration then ensues.  With
no diffusion, large gaps remain pristine. Although slow, type I
torques can cause radial drift on very long timescales, fast migration
cannot begin until the moon is within a corotation radius of either
edge of the gap.  Therefore, eliminating particle diffusion in a large
gap effectively eliminates fast migration.

\subsubsection{Fast migration in wide gaps: encroaching ring edges}


To provide more details on how a moon within a large gap begins fast 
migration, we consider another example. The simulation begins with a
small moon, $\rhill$ = 4.9~km, in a perfectly circular orbit within
a large gap of half-width $\Delta a$ = 20~km $\approx 4 \rhill$. The
gap is centered within a larger ring of particles with properties 
appropriate for the A ring, equations~(\ref{eq:Sigmafid}--\ref{eq:nuradfid}). 
As in other calculations, ring particles have a diffusion time of 10 yr.
The simulation spanned more than 50~years, or 25,000~orbits. A sequence 
of snap shots shows how gap edges are sculpted and how particle orbits 
diffuse close enough to the satellite to trigger fast migration. 

Figure~\ref{fig:daphnis2x96} illustrates how ring material slowly diffuses
into a wide gap and eventually leaks onto orbits that interact strongly 
with the small moon. Before the moon starts to migrate, very little of 
this material enters the moon's corotation zone. Instead, slight imbalances
draw the moon slightly closer to one gap edge than the other. This process
is slow. In the top two panels of Figure~\ref{fig:daphnis2x96}, the time
difference is roughly 40 yr.  Only slight changes are visible.  The gap 
is still well defined. The wakes created by the moon are clear.

Once changes begin, they are rapid.  In Figure~\ref{fig:daphnis2x96}, the 
lower two panels show the scene only a few years later. Orbit crossing is 
rampant. As the moon drives into one side of the gap, it pulls material 
across, opening a new gap ahead of it and filling in the space behind it.  
This new, smaller and less well-defined gap comoves with the satellite 
\citep[see][and the associated animation in a much broader disk]{bk11b}.  
Although the original gap remains behind, viscous diffusion slowly fills 
it in.

Tests with moons of other sizes in gaps of varying widths yield similar
results. Diffusion tries to fill the gap; scattering tries to empty it.
Eventually, there is an imbalance which pushes the moon towards one edge
of the gap. The accelerating imbalance between scattering and viscous 
diffusion allows the moon to draw more and more material from the gap and
to move closer and closer to the edge of the gap. Once the imbalance 
reaches a critical level, fast migration begins.

The starting conditions for the results shown in Figure~\ref{fig:daphnis2x96} 
are {\it almost} a realistic representation of Daphnis, the A ring and 
the Keeler gap. The single difference is the initial orbit of the moon.
In the simulation, the initial orbit has $e = i = 0$. The real orbit
is slightly eccentric and inclined to the ring plane. As we show in \S5,
the time required for a moon to begin fast migration depends on the
initial $e$ and $i$. However, the overall outcome is identical: the moon
always begins fast migration.

\subsubsection{In situ formation and clearing the corotation zone}

When it grows by coagulation/accretion, a satellite forms within a
dense disk of particles. With a full corotation zone, this satellite
cannot migrate in the fast mode. Previous numerical experiments 
\citep[e.g.,][]{kir09} demonstrate that a satellite can gravitationally 
scatter away enough of the corotating material to trigger fast migration 
in a particle disk.  Here, we illustrate the same phenomenon for a small 
moon in Saturn's A ring.  Figure~\ref{fig:mhorse} shows results for a 
simulated moon with $\rhill = 14$~km in the A ring. The figure tracks 
the orbital distance and the amount of material in the corotation zone 
as a function of time.  After the mass in the corotation zone falls to 
about a third of its starting value, fast migration starts. The mass of 
corotating material then peaks up as the moon plows through fresh material 
during its inward spiral through the disk.

Figure~\ref{fig:mhorse} shows that a small moon rapidly clears its corotation
zone \citep[for other examples of this phenomenon, see also][]{raf01,kir09}.  
Within 10--20 yr, gravitational scattering removes roughly half the mass of 
the corotation zone. During this period, the moon slowly moves radially inward 
and outward in response to perturbations from swarm particles in the corotation 
zone. After $\sim$ 100 yr, the moon has removed enough material from the 
corotation zone to begin fast migration. After migrating inward roughly 200 km, 
the moon's progress is halted as it encounters the edge of the disk. 

Our simulations demonstrate that this behavior is independent of the
general background perturbations from other small moons and larger
distant ones orbiting Saturn.  Comparison of calculations without
distant moons (Fig.~\ref{fig:mhorse}, black curve) and with a full
complement\footnote{In these simulations, a `full complement' of moons
  includes the ring moons Pan, Daphnis, Atlas, Prometheus, Pandora,
  Epimetheus, and Janus and the larger, more distant moons Mimas,
  Enceladus, Tethys, Dione, Rhea, Titan, and Iapetus.} of moons
(Fig.~\ref{fig:mhorse}, grey curve) shows modest differences resulting
from stochastic variations in the orbits of swarm particles but no
large variation in the timing or rate of migration.  Results for a
large suite of simulations yield the same conclusion: fast migration
of satellites with $\rgap < \rhill < \rfast$ is a robust phenomenon in
the A ring.  Thus, the general lack of small moons in the ring is
consistent with migration theory \citep[see also][]{gt82}.

\subsubsection{Orbital settling into a ring}

The onset of fast migration depends on the initial orbital parameters
of a satellite.  When a small moon is scattered into an otherwise 
unperturbed region of a ring, it will probably have a large orbital 
eccentricity or inclination relative to the ring particles. The moon
then tries to clear a region much larger than its corotation zone.
When $e \gtrsim 10^{-5}$, the radial excursion of the moon exceeds the 
radius of its corotation zone. When $i \gtrsim 10^{-7}$, the moon spends 
most of its time outside the ring and cannot completely clear out material 
along its orbit. However, gravitational interactions with the ring particles 
damp $e$ and $i$ \citep[e.g.,][]{gt81,bor84,war88,hahn2007,hahn2008}. 
Once $e$ and $i$ are small, the moon can clear its corotation zone and begin 
fast migration.

To derive the damping time for a small moon with an inclined orbit, we
perform a set of calculations with various initial $e$ and $i$
relative to swarm particles in the A ring.  Figure~\ref{fig:capture}
illustrates the rapid damping of $e$ and $i$ experienced by a small moon
in a typical calculation.  For moons capable of undergoing fast
migration, the damping timescale is well within 1--2 decades.  The
inclination damps smoothly and reaches $i < 10^{-6}$ in 10--20 yr.
After moons reach $e < 10^{-4}$, the drop in $e$ is more
gradual and includes additional stochastic variations due to
perturbations in the surface density of swarm particles.

Once a small moon achieves a nearly circular orbit, the timescale to 
damp the orbit further is much longer. When the inclination satisfies
$\rhill \lae a i$, the moon can begin to clear its corotation zone. 
The timescale to reach this inclination is roughly the viscous time
to fill the gap formed when the moon had much larger $e$ and $i$. 
Figure~\ref{fig:daphnist100} provides an illustration. A small moon, with 
$\rhill = 6.5$~km, experiences a slow damping followed by a transition 
to fast migration.

Simulations of the orbital evolution of small moons with $\rgap <
\rhill < \rfast$ and a broad range of starting $e$ and $i$ yield
similar results. As the moons try to clear a gap much larger than their
corotation zones, their $e$ and $i$ damp. While $e$ and $i$ damp,
viscous spreading fills in the gap outside a moon's corotation
zone. After $i$ reaches a threshold, fast migration begins.

The vertical scale height of the ring sets the threshold for fast
migration of a small moon with an inclined orbit relative to the ring. 
The condition for fast migration is $i_{\rm fast} < \rhill / a$. When 
$i \gtrsim \rhill / a$, moons cannot completely clear their corotation 
zones before viscous spreading fills them in.  After damping reduces $i$ 
to the threshold $i_{\rm fast}$, fast migration begins.

We conclude that fast migration may be a common, if not ubiquitous,
process in a disk similar in structure to Saturn's A ring.  The
migration time is extremely short compared to the $\sim$ 0.1--1 Gyr
lifetime of the ring \citep[e.g.,][]{espo2010}. Thus, (i) all moons 
with $\rhill \gtrsim \rgap$ are recent additions to the ring or 
(ii) physical structures within the ring prevent migration. To 
address these possibilities, we next consider how density 
perturbations within the ring impact migration.  In \S5, we examine 
the possibility that the moons are young.

\subsection{Fast versus stochastic migration}

The A~ring is dynamically active, with a variety of long-lived and
transient physical structures. Many features are associated with
spiral density waves \citep[e.g.,][]{cuz1981} or bending waves
\citep[e.g.,][]{shu1983}.  In addition to the surface density
enhancements immediately surrounding moonlets (the propellers),
nearby moonlets or small moons generate gravitational wakes in the
rings \citep{show1986}. Additional variations of optical depth in the
A ring result from self-gravity wakes \citep{hed07}, which vary in
structure across orbital resonances with Saturn's moons.

Although simulating the full range of structure present in the A ring 
is beyond the current capabilities of \codename, we can establish how
a variety of density perturbations impact fast migration in the A ring. 
We begin with simulations designed to illustrate how clumpiness in the 
ring impacts migration.

Figure~\ref{fig:massrat} shows a suite of simulations of a moon
with $\rhill$ = 14~km at an orbital distance of 130,000~km embedded 
in rings with a fixed surface density and with swarm particles of 
various sizes.  The swarm particles in each simulation all have the 
same mass, but from run to run, that mass ranges from 0.2\% to 10\% of 
the moon mass.  In all simulations, the corotation zone is initially 
free of disk particles in an annulus of $3\rhill$. Fast migration 
begins nearly immediately. Without this initial clearing of the
corotation zone, there is a time lag $\sim \tclear$ in the onset 
of fast migration \citep[see Fig. 5 and ][]{kir09}. 

These simulations demonstrate that density perturbations can change 
migration rates substantially.  As the mass of a swarm particle 
increases, migration through the ring becomes more and more erratic 
(Fig.~\ref{fig:massrat}).  When swarm particles are 10\% of the
satellite mass, migration begins to resemble a random walk process, 
as envisioned in the stochastic migration scenario of 
\citet[][see also Tiscareno 2012]{cri10}.

For a more detailed illustration of discreteness effects in migration,
we consider the migration of a small moonlet with $\rhill$ = 700~m. As
in Figure~\ref{fig:mhorse}, the calculations include all of Saturn's
major moons (e.g., Titan and Rhea) and smaller moons near the A ring
(e.g., Prometheus and Pandora).  To enable computational feasibility,
we reduce the number of swarm particles by decreasing the surface
density of the A ring by a factor of 100, setting the mass of swarm
particles at 1--2\% of the moonlet mass, and clearing the corotation
zone of swarm particles.  We also turn off viscous diffusion. Without
viscosity, particles with Hill radii comparable to the propeller
moonlets ($\rhill \lae 1$~km) can maintain a fairly empty corotation 
zone once they start migrating.

Figure~\ref{fig:moonletmigrate} shows typical results of these
simulations.  The noisy horizontal curve in light grey measures the
fluctuations in semimajor axis -- as measured in a reference frame
centered on Saturn -- of an isolated moonlet orbiting the planet but
not migrating through the disk.  These fluctuations result from the
constant gravitational jostling from distant and nearby moons.  When
the moonlet is allowed to migrate (as in the black and dark grey
curves, which have particles that are 1\% and 2\% of the moonlet mass, 
respectively), its semimajor axis shifts in a stepwise fashion which tracks
the theoretical prediction (dashed curve). Over a one year time frame,
the total migration is comparable to the variation in $a$ from
jostling by distant moons.

These two simulations illustrate the difficulty of measuring migration rates
in Saturn's rings and other complex systems. Gravitational jostling of orbits
by distant moons is comparable to the expected radial excursions of small 
moonlets comparable in size to Bl\'{e}riot and other propellers in the rings.  
Interactions with larger ring particles can hinder or reverse migration of
a larger moon on short timescales \citep[e.g.,][]{tis12}.  In all cases, 
measuring robust migration rates requires long time baselines and stable 
reference frames.

\subsection{Migration in the presence of distant moons: resonances}

In addition to propellers and wakes, Saturn's moons create a variety of
structure within the A ring. The moons Pan (Encke gap) and Daphnis (Keeler gap)
lie within distinct low density gaps within the rings.  A 7:6 resonance 
with the coorbiting moons Janus and Epimetheus is responsible for the 
outer edge of the A ring \citep[e.g.,][]{lis85}. The ``ring moons'' Atlas,
Pandora, and Prometheus just outside the ring also produce structure in 
the ring.

To illustrate the rich set of other resonances in the A ring
\citep[see also][]{lis82},
Figure~\ref{fig:fullring} shows the result of a simulation of 10,000 
tracer particles orbiting Saturn along with the major moons (Mimas, 
Enceladus, Tethys, Dione, Rhea, Titan, and Iapetus) and several 
smaller moons much closer to the ring (Pan, Daphnis, Atlas, Prometheus, 
Pandora, Janus and Epimetheus). Ring particles have large maxima in $e$ 
around Pan (the Encke gap) and at the outer edge of the ring (the 7:6 
resonance of Janus and Epimetheus).  Other rises in $e$ occur at the
orbits of Atlas and Prometheus -- which lie beyond the outer edge of 
the ring in regions of very low surface density -- and at 
many other labeled resonances. Several of these resonances lie close 
to but are not obviously coincident with the orbit of Daphnis within
the Keeler gap. Curiously, a large ensemble of propeller moonlets lie
between the 5:4 resonance of Janus/Epimetheus and the 5:3 resonance with
Mimas.

To explore the impact of orbital resonances on migration, we consider
whether small moons can pass through the 5:4 resonance with Janus and
Epimetheus. We set up an ensemble of moons with $\rhill$ = 3--15 km on
either side of the resonance near $a = 130,500$~km. Because small
moons can migrate inward or outward, some moons attempt to migrate
across the resonance. Others try to migrate away from the
resonance. In all cases, Janus/Epimetheus stir up the eccentricities
of ring particles close to the resonance. Large vertical scale height
and large eccentricity are barriers to fast migration. Thus, we expect
the migration of smaller moons to stall close to resonance.

The results in Figure~\ref{fig:avtjanepi} confirm our expectations.
Resonant interactions with Janus and Epimetheus stir up enough ring 
material to create an obstacle for smaller moons. Moons with 
$\rhill = 3.5$~km do not cross the ``barrier'' set up by the 5:4 
resonance. However, moons with $\rhill$ = 14 km migrate 
right through the resonance.

To understand this resonant barrier to migration, we examine the
eccentricity of ring particles close to and far away from the
resonance.  At resonance, Janus and Epimetheus can stir ring particles
to fairly high eccentricity ($\eice \sim 10^{-5}$; see
Fig.~\ref{fig:fullring}).  Away from resonance, $\eice$ is a factor of
$\gtrsim$ 10 smaller.  At a distance of 130,000~km from Saturn, these
ring particles have radial excursions of a few kilometers. When they
are at resonance, small moons or moonlets with Hill radii smaller than
a few km are thus embedded in a dynamically hot ring, which
dramatically reduces their migration rate
\citep[][eq.~21]{bk11a}. Moons with Hill radii larger than a few km
lie in a dynamically cold ring whether or not they are within the
resonance. These larger satellites migrate freely.

The value of $\rhill$ that separates these two modes of migration
through a resonance depends on several factors. With other properties
of the rings held fixed, stronger (weaker) resonances produce stronger 
(weaker) eccentricity enhancements, creating barriers for larger 
(smaller) moons.  If the strength of the resonance is held fixed,
stronger (weaker) damping or viscous spreading weakens (strengthens)
eccentricity enhancements, enabling (preventing) migration. Thus,
understanding barriers to migration requires a thorough understanding
of the physical processes that impact the eccentricity of ring particles.

\subsection{Migration in the presence of distant moons: stirring}
\label{subsect:distmoons}

Although distant moons outside of resonance weakly stir the eccentricity 
and inclination of ring particles, this stirring can have a major impact
on the structure of the rings.  Using the impulse approximation, 
\citet{wei89} showed that distant moons give a particle a small kick
in $e$ and $i$ every synodic period. For an ensemble of distant moons,
he derived a stirring rate
\begin{eqnarray}
\label{eq:estir}
\frac{d e^2}{dt} & \sim & 
 \sum_{n} \frac{m_n^2}{M^2}\frac{a^3}{|a-a_n|^3}\ \frac{1}{T}, 
\\
\label{eq:istir}
\frac{d i^2}{dt} & \sim & 
 \sum_{n} \frac{m_n^2}{M^2}\frac{(i^2+i_n^2) a^5}{|a-a_n|^5}
\ \frac{1}{T}, 
\end{eqnarray}
where $m_n$, $a_n$ and $i_n$ are the mass, orbital distance and inclination 
of the $n^{\rm th}$ moon. These expressions are valid for objects on nearly 
circular orbits.  

For the Saturn system, this approach is valid for moons with separations
from the ring, $\delta a$, much smaller than the semimajor axis of ring
particles.  Thus, these equations apply to moons as far as Janus and 
Epimetheus (and possibly Mimas), but are invalid for more distant moons 
(e.g., Titan). With an orbital distance of $1.2\times 10^6$~km, Titan
tugs more on the barycenter of the entire saturnian system instead of 
stirring the rings.

Our calculations yield results in rough agreement with the analytic theory.
In several simulations for Figure~\ref{fig:fullring}, the eccentricity of
ring particles grows at a rate $de^2/dt \sim 10^{-11}$~yr$^{-1}$, within a
factor of two of the prediction from eq. (\ref{eq:estir}).  With typical 
$ e \lesssim 10^{-6}$ for ring particles, the growth time for stirring by
distant moons is a fraction of a year, roughly equivalent to the damping time.

To compare stirring by distant moons with collisional damping in more detail, 
we consider the energy involved in both mechanisms. Stirring by distant moons 
results in a kinetic energy input per unit mass of 
${\cal P}_{\rm stir} \sim \vkep^2 de^2/dt \sim 2\times 10^{-7}$~dyne~g$^{-1}$.  
The power per unit mass dissipated by collisions is
\begin{eqnarray}
\label{eq:Pcoll}
{\cal P}_{\rm coll} & \sim & \frac{\Sigma}{\mice h} \vice \pi \rice^2
\floss \mice \vice^2 \\
\ & \sim & 6\times 10^{-7} \ {\mbox{dyne/g}},
\end{eqnarray}
where $\mice$ is the characteristic mass of the ring particles.
To derive this estimate, we adopt ring particles with $\rice$ = 1~m
and $\rhoice$ = 1~g~cm$^{-3}$.  The rough equality of these rates suggests 
that the A ring's velocity dispersion and thickness are supported by 
stirring from distant moons.  A precise balance between the two energies
requires modest changes in the size/density of the ring particles.

To examine the impact of stirring by distant moons in more detail, we 
consider simulations of a single ring particle interacting with Mimas,
the innermost massive moon of Saturn. Each calculation begins with a
ring particle orbiting at $a \approx$ 136,500~km, with
$e \approx 3 \times 10^{-5}$ and $i \approx 6 \times 10^{-5}$ relative
to the ring plane\footnote{These elements are similar to those of Daphnis.}.
Mimas begins at a random phase in its orbit, which is tilted by roughly
1\degree\ relative to the ring plane. 

Figure~\ref{fig:stir} shows the results. In a basic set of simulations,
Mimas rapidly stirs the $e$ and $i$ of the ring particle (left panels
of figure). The growth rate of $e$ follows the \citet{wei89} estimate 
closely. The growth rate of $i$ is several times faster than predicted. 
In these simulations, a distant perturber (Mimas) continuously forces
the alignment of angular momentum vectors between Mimas and the ring 
particle. Thus, the impulse approximation fails and $i$ grows rapidly.
Although adding collisional damping slows the growth of $e$ and $i$
(right panels), $e$ and $i$ still grow rapidly. 

When we introduce the oblateness of Saturn into these simulations, the
growth of $e$ and $i$ stalls. In the gravitational field of an oblate
Saturn, the ring particle precesses rapidly. Because the precession
time is short compared to Mimas' orbital period, Mimas cannot force
alignment of angular momenta. Thus, a single ring particle maintains
a constant $e$ and $i$ (Fig.~\ref{fig:stir}, left panels, light grey
lines). Reducing the precession rate by a factor of 100 enables Mimas
some impact on the ring particle; $e$ and $i$ then vary periodically
(Fig.~\ref{fig:stir}, left panels, dark grey lines).

Including damping and oblateness into the calculations produces a
different equilibrium between stirring and damping. With a reduced 
precession rate (Fig.~\ref{fig:stir}, right panels, dark grey lines),
$e$ and $i$ oscillate about a slowly declining median level. With 
a normal precession rate around an oblate Saturn, $e$ and $i$ decline
monotonically (Fig.~\ref{fig:stir}, right panels, light grey lines).
At late times, $t \gg 10^2$ yr, $e$ and $i$ settle into an equilibrium
with $e \approx i \approx 10^{-7} - 10^{-6}$ set by the damping rate. 
Larger damping results in a smaller equilibrium $e$ and $i$.

This suite of simulations demonstrates that a balance between damping
and stirring by distant moons sets an equilibrium $e$ and $i$ for ring
particles. Throughout the ring, the damping time grows in regions of
reduced surface density (eq. [\ref{eq:tdamp}]). Thus, stirring by
distant moons is more important in regions of low surface density.
Because migration slows in regions with large $e$ and $i$, stirring
by distant moons can inhibit migration more easily in regions of low
surface density. 

%

\section{Daphnis}
\label{sect:daphnis}

If the migration theory outlined in \S\ref{sect:theory} is correct,
the A ring should contain no moon or moonlet larger that a few
kilometers.  Larger satellites migrate to the ring edges on timescales
much shorter than the ring lifetime. Although some smaller moonlets --
like the propellers -- can migrate slowly, our simulations show that
they cannot pass through modest resonances.  Aside from Daphnis, the
ring has no large moonlets or small moons capable of migrating through
the resonances shown in Figure~\ref{fig:fullring}.  Because Daphnis
should migrate so swiftly, understanding the origin and the behavior
of Daphnis is essential to migration theory.

With an orbital distance of $136,505$~km from Saturn, Daphnis is embedded 
in the $\sim$ 20~km wide Keeler gap \citep{wei09}. It has a modest
eccentricity $e \approx 3\times 10^{-5}$ and inclination $i \approx 
6.3\times 10^{-5}$ \citep{jac08}.  Relative to a circular orbit in 
the ring plane, the radial and vertical excursions of Daphnis have 
amplitudes of roughly 10~km, about twice the moon's Hill radius of 
4.9~km \citep{tho10}.

Daphnis is a mystery.  Despite having a Hill radius within the range 
identified for fast migration, its radial drift is negligible. Using
the measured viscosity of the A ring \citep[e.g.,][]{tis07,col09},
the timescale for ring particles to circularize Daphnis' orbit is
$\sim$1500~yr (eq.~[\ref{eq:tfill}]; see also Hahn 2008). Reducing the 
inclination to zero follows soon thereafter \citep[see also][]{hahn2007}, 
allowing fast migration to commence. With an expected total lifetime in 
the interior of the A ring, $\sim$ 2500~yr, much shorter than the age 
of the rings, migration models are closely linked to models for the 
origin of Daphnis.

\subsection{Origins}

There are three possible origins for Daphnis. If this small moon is a
recent addition, it might be a fragment produced during the impact
of a comet or an asteroid with the ring system. Alternatively, it 
might have formed {\it in situ}. We reject the idea that the moon 
results from a three (or more) body interaction with Saturn's major 
moons and outer irregular satellites.  The escape velocities of 
Saturn's moons are too low to scatter the moon into the rings.  
If the current orbit of Daphnis is stable for timescales much much
longer than the $\sim$ 1000~yr migration time, Daphnis is a much 
older resident of the rings, perhaps a remnant of the tidal event 
that produced the ring system. In this scenario, some set of external
processes prevents Daphnis from migrating.  To judge which of these
possibilities is more likely, we consider each in more detail.

Although there is a low probability that Daphnis is an impact fragment
\citep[e.g.,][]{cuz90}, this scenario is easy to test with observations.
The current orbital eccentricity and inclination are appropriate for a
small moon settling into the ring (Fig.~\ref{fig:capture}). The numerical 
simulations demonstrate that $e$ and $i$ damp to zero on timescales of 
100--1000 yr \citep[see also][]{hahn2007,hahn2008}.  For Daphnis' orbit 
in the A ring, we expect the maximum altitude above the ring to decrease 
by 10--100 meters per year.  Although there are no current observational 
limits on this rate, a direct measurement would constrain this 
`recent external origin' scenario.

Forming Daphnis {\it in situ} is more likely. \citet{cha10,cha11} show 
that many of the inner moons of Saturn may have grown from ring 
material. Although Daphnis lies inside the orbits of the moons
considered in \citet{cha10}, it is reasonably close to the outer edge
of the A~ring.  If Atlas, Prometheus, and other larger moons formed
just outside the ring, it is reasonable to suppose Daphnis formed with
them and migrated inwards to its present location.

Despite the attraction of this idea, Daphnis' current orbit still 
requires special timing. The size of the Keeler gap surrounding 
Daphnis' orbit is roughly the correct size for a gap produced by a 
moon like Daphnis.  If Daphnis cleared this gap, then it should 
migrate on short timescales (Fig.~\ref{fig:mhorse}).  Thus, for 
{\it in situ} formation to be viable, it probably happened relatively 
recently.

Although this conclusion seems at odds with the \citet{cha10} simulations,
observations can test this `recent in situ origin' scenario. Because 
changes in altitude above the rings and the radial motion of Daphnis 
are detectable, direct measurements can establish whether the orbit 
of Daphnis is damping vertically or migrating radially.

Because these scenarios seem so unlikely, we consider other processes 
that might prevent rapid migration of Daphnis. As shown in the simulations 
for Figure~\ref{fig:avtjanepi}, distant moons may establish a resonant 
barrier which prevents radial migration. In a cursory search for candidates, 
the Prometheus 32:31 and Pandora 18:17 resonances seem promising.  Although 
the chaotic behavior of the orbits of Saturn's inner moons complicates 
measuring the strength of resonances and their impact on Daphnis, our 
simulations demonstrate that resonances can stop the radial migration of
moons similar to Daphnis.

Gravitational stirring by distant moons may also prevent migration
(Fig.~\ref{fig:stir}).  Several of the larger moons -- notably Mimas
and Tethys -- have orbits with inclinations $i \approx$
1\degree\ relative to the ring plane. Mimas and Tethys are also locked
in a 2:4 mean motion resonance
\citep[e.g.,][]{all1969,vie1992,champ1999}. Stirring by Mimas and Tethys 
counter viscous damping by the rings. From equations
(\ref{eq:estir})--(\ref{eq:istir}), we estimate that Mimas doubles the
eccentricity of Daphnis in less than a decade. With its larger mass,
Tethys can double Daphnis' eccentricity in 1.5~yr.  Because these
timescales are comparable to the damping time, stirring plausibly
slows down migration.


\subsection{Simulations}

To test the two possibilities to halt Daphnis' migration in more detail, 
we turn to a set of numerical calculations.  Currently, we cannot 
simulate the full set of physical processes which impact Daphnis' orbit.
Our goal is to show that resonances and stirring by more distant moons
are plausible mechanisms to prevent migration in Daphnis.

We begin with a suite of test runs for a single moon with the Hill
radius of Daphnis embedded within the Keeler gap. Starting from Daphnis'
current $e$ and $i$, the moon rapidly aligns with the ring 
\citep{hahn2007, hahn2008}.  With a roughly linear decrease in the 
vertical excursion of Daphnis of about 50~m~yr$^{-1}$, Daphnis' maximum 
height above the ring plane drops from $\sim$ 10~km to much smaller than 
1~km in roughly 200 yr (Fig.~\ref{fig:daphnis}). Once Daphnis has a 
vertical amplitude smaller than the scale height of the ring, it begins 
fast migration.

For the next set of simulations, we consider a Daphnis-sized moon
in the 32:31 resonance with Prometheus. The resonance pumps up the 
eccentricity and then maintains a roughly sinusoidal variation in $e$ 
for $\sim$ 10~yr. As $e$ oscillates, the resonance pushes the moon 
outside the gap. Once the moon lies within pristine ring material, 
damping dominates stirring by the single outer moon. Both $e$ and $i$ 
decline dramatically. When $i$ falls below the threshold 
$i_{\rm fast} < \rhill / a$, the moon commences fast migration 
(Fig.~\ref{fig:daphnis}, light grey line). 

Moving Daphnis from the Prometheus 32:31 resonance to the Pandora 18:17 
resonance leads to similar behavior on longer timescales. Within the Pandora 
resonance, Daphnis' orbit remains inside the Keeler gap for $\sim 20$~yr.
The resonance then moves Daphnis out of the gap into the ring. Although
it takes awhile for the ring to damp Daphnis' inclination, $i$ eventually
drops below $i_{\rm fast}$. Daphnis then begins fast migration.

When we move Daphnis' orbit to an off-resonance location, its orbit
becomes more stable, at least on decade-long time scales
(Fig.~\ref{fig:daphnis}, light grey line). In this set of
calculations, Daphnis settles into the ring on a time scale comparable
to calculations without Prometheus and Pandora, or perhaps even
faster.  There is no clear sign that stirring from these small,
nearby moons can prevent settling. However the moons do cause some
small differences in the evolution of orbital elements, which
is an encouraging sign: during the first 5000 simulated orbits (7.5~yr),
Daphnis' inclination shows very little damping.

These simulations demonstrate the ability of interactions with the 
distant moons to affect the orbital parameters of Daphnis-sized 
moons on very short timescales.  Finding plausible equilibria for 
Daphnis within the Keeler gap requires a detailed parameter search to 
identify the best combination of initial $(a, e, i)$ for the satellite 
and the outer moons, and $(\Sigma, \nurad, \tdamp)$ for the ring.  Although 
several test simulations suggest it is possible to find equilibria stable 
over million year timescales, the computational effort is significant 
and beyond the scope of this paper.

In a final set of simulations, we consider the reaction of Daphnis to
non-resonant interactions with all of the distant moons. The
simulations with Mimas in \S4.5 show that a balance between damping
and stirring from a single moon can maintain a large inclination, with
$i \gtrsim i_{fast}$. Stirring by Prometheus and Pandora can reduce
settling rates, at least over several thousand orbits.  Thus,
stirring from all the outer moons could maintain a large inclination,
probably with some jitter due to the constantly changing gravitational
field along Daphnis' orbit.

As in Figure~\ref{fig:fullring}, these simulations include all of the 
major moons and the ring moons.  The coordinates of the moons and 
Daphnis relative to Saturn come from the JPL Horizons systems 
\citep{gio96} 00:00:00~GMT, 2012-June-22; we adjusted the ring plane 
to enable starting conditions for Daphnis similar to those in
Figure~\ref{fig:daphnis}. The calculations include the A-ring with an 
empty Keeler gap, damping and diffusion of ring particles, and Saturn's
oblateness.

Figure~\ref{fig:daphnisall} shows the results of one simulation. As in
Figure~\ref{fig:daphnis}, the thin black curve shows the decay of $e$
and $i$ of a lone Daphnis orbiting within the Keeler gap and
interacting only with ring particles. Orbital interactions with the
full ensemble of Saturn's moons yields relatively rapid, small
amplitude fluctuations in $a$ and $e$. This jitter is similar in
amplitude to the jitter of any small moon in the ring (see
Fig.~\ref{fig:moonletmigrate}). On much longer timescales of a few
months to half a year, Saturn's moons produce 0.5~km oscillations in
the $z_{\rm max}$ of the moon. On decade timescales,
$<$$z_{max}$$>$, the median height of the moon above the ring plane
is roughly constant. Thus, stirring by Saturn's moons appears capable 
of maintaining the large inclination of Daphnis relative to the A~ring.

The predicted oscillations of Daphnis are sensitive to several
characteristics of our calculations. In our approach to ring dynamics,
ring particles have no self gravity and therefore do not react to
changes in local surface density. Furthermore, the ring plane itself is fixed
and does not evolve in response to torques.
%
%
Despite these limitations, the calculations 
make a robust prediction: as the mass of a small moon grows and as the
surface density of ring particles around the satellite declines, the
distant moons of Saturn produce an oscillatory behavior in the 
height of the small moon above the ring plane.  These oscillations
can prevent small moons like Daphnis from migrating through the ring.

The predicted quasi-periodic oscillations in the altitude of Daphnis 
above the ring plane are detectable. The predicted amplitude of 
200--400~m is a significant fraction of the typical maximum altitude 
of roughly 9~km. The quasi-period of a few months yields a rate 
$\dot{z}_{max} \approx$ 0.4--0.8~km~yr$^{-1}$, a few times larger
than the rates of motion derived for the propeller moon Bl\'{e}riot 
\citep{tis10}. Thus, accurate observations of Daphnis' inclination 
or maximum height above the ring plane would test our interpretation.

\subsection{Summary}

Our results suggest several plausible origins for Daphnis.  A `young' 
Daphnis can result from an impact fragment or from growth beyond the
outer edge of the A ring.  In both of these recent origin scenarios, 
ring material on either side of the Keeler gap should damp Daphnis'
orbit on short, $\sim$ 1000 yr, timescales. Interactions between 
Daphnis and the rings should also induce fast migration.

An `old' Daphnis has been resident in the rings for timescales exceeding
$\sim$ 10,000 yr. Because external processes prevent Daphnis from migrating, 
Daphnis' residence time in the ring could approach the ring lifetime of
$\gtrsim$ 0.1--1~Gyr. Our simulations show that orbital resonances with distant
moons can impact Daphnis' orbit. Identifying the combination of physical
properties for Daphnis, the rings, and the distant moons needed to maintain 
Daphnis in its current orbit requires a comprehensive suite of simulations 
that is beyond the scope of this paper.

Our simulations also demonstrate that orbital interactions between 
Daphnis and distant moons (e.g., Mimas and Tethys) outside of any 
resonance can impact Daphnis' orbit. Maintaining Daphnis in its current 
orbit requires a fine balance between stirring by the distant moon(s)
and damping by ring particles outside the Keeler gap. 

\subsection{Observational Tests}

Observations can test all of these scenarios. If Daphnis is young, we
predict detectable, monotonic changes in the orbital parameters 
$(a, e, i)$.  If Daphnis is old, changes in $(a, e, i)$ depend on
the importance of resonances. Outside resonance, we expect modest 
changes in phase with the motions of distant moons such as Mimas. 
Within resonance, the behavior depends on the strength of the
resonance relative to damping by the ring.  Although measuring orbital 
parameters in a chaotic reference frame is challenging, direct 
measurement of Daphnis' altitude above the ring plane and position 
in the Keeler gap provide equivalent tests of our predictions.  


\section{Conclusions}

Saturn's A ring is rich in structure, with an amazing, complex variety 
of transient and long-lived features \citep[e.g.,][]{espo2010}.  The 
smallest ring particles lie in a matrix of self-gravity wakes, which 
form coherent features on scales smaller than 100~m \citep{hed07}. 
The 150 or so moonlets embedded in the ring have physical radii 
between 50~m and 250~m, and are located in three $\sim 1000$~km 
annuli \citep{tis08}. The small moons Pan and Daphnis orbit within 
distinct gaps with very low surface density. Together, these moons, 
moonlets, and smaller particles provide an excellent testbed for migration 
theory \citep[see also][and references therein]{cri10,pan10,rei10,sal10a,tis12}.

As outlined in Figures~\ref{fig:migratesketch} and \ref{fig:migrateAring},
theory predicts several distinct modes of migration through the rings.
In the very viscous A~ring, moonlets with $\rhill \lesssim$ 0.1~km
migrate on timescales much longer than the age of the solar system. 
Large moons with $\rhill \gtrsim$ 20--30~km also migrate very
slowly. Between these limits, migration timescales are shorter than
the ring lifetime. For $\rhill \approx$ 2--20~km, the timescales 
are comparable to a human lifetime.

In our simulations of Saturn's A ring, fast migration of small moons with
$\rhill \approx$ 2--20~km is ubiquitous. When a small moon lies within a 
gap, it eventually pulls enough material from the gap edges to begin fast 
migration (Figs.~\ref{fig:ringxx}--\ref{fig:daphnis2x96}). When a small 
moon is embedded within a continuous ring, it eventually clears enough 
material from its corotation zone to begin fast migration 
(Fig.~\ref{fig:mhorse}).

In a pristine ring of small particles, numerical simulations show that
moonlets and small moons migrate at the rates predicted by analytic
theory (Figs.~\ref{fig:massrat} and \ref{fig:moonletmigrate}).  In
numerical simulations that include damping, gravitational interactions
between small moons and ring particles, and viscous spreading, the
moons clear their corotation zones on timescales $\tclear$. When the
viscous timescale is longer than $\tclear$, these moons open a gap in
the ring and begin to migrate. As long as $\rhill \lesssim \rfast$,
migration is fast. Otherwise, migration is slow.

In Saturn's rings, damping and viscous spreading smooth out the wakes
left behind by migrating moons. Thus, multiple moons can easily
migrate rapidly through the rings. This situation contrasts with
protoplanetary disks, where stirring by migrating protoplanets 
overcomes damping by smaller planetesimals \citep{bk11b}.


Several physical processes within the rings limit or halt migration.
Interactions between small moons and (i) very large ring particles or
(ii) local enhancements in ring surface density lead to stochastic
migration, often reducing migration rates significantly
\citep{cri10,rei10,tis12}.  Migrating moonlets and very small moons
cannot pass the `resonant barriers' formed by moons outside the A~ring
(\S4.4). Pairs of resonances can `corral' these objects and prevent
them from migrating in or out through the ring
(Fig.~\ref{fig:fullring}). Stirring by distant moons outside resonance
can also overcome damping and prevent migration (\S4.5).

These results lead to several clear conclusions regarding migration in
the A~ring.

\begin{itemize}

\item {\bf Propeller moonlet migration.}  With Hill radii less than
  $\rgap \approx 2$~km \citep{tis08}, the propeller moonlets should
  experience slow, embedded migration with radial drift rates below
  10~m~yr$^{-1}$ (Fig. \ref{fig:migrateAring}).  Stochastic migration
  in a self-gravitating ring \citep{rei10,cri10}, interactions with
  surface density variations in the ring \citep{tis12}, and
  oscillatory motion from interactions with corotating material
  \citep{pan10,pan12a} have similar low drift rates.  For most
  moonlets, measurements of radial drift limit the rate to below
  0.1~km~yr$^{-1}$.  At least one larger moonlet, Bl\'{e}riot, shows
  non-Keplerian motion \citep{tis10}, with a drift rate that varies in
  time. Our new results suggest that the radial drift of Bl\'{e}riot
  and other moonlets can be confined by orbital resonances. Indeed,
  most propellers appear to be confined to the `corral' between the
  Mimas 5:3 resonance and the Janus/Epimetheus 5:4 resonance
  (Fig.~\ref{fig:fullring}).

\item {\bf The propeller moonlet size distribution.}  Moonlets have
  a range of sizes up to about $\rhill \sim 1.5$~km \citep{tis08,tis10}. 
  Migration theory predicts an upper limit of $\rgap \approx 2$~km.
  Larger objects migrate out of the ring on timescales of 100--1000~yr.
  The good agreement between $\rgap$ and the maximum size of propellers
  is a strong confirmation of migration theory.

\item{\bf Non-Keplerian motion of moonlets.} Moonlets orbit Saturn in 
  a gravitational field which is constantly modified by much more massive 
  moons orbiting at larger distance. The constant jostling of moonlets
  by distant moons results in chaotic radial variations in distance from
  Saturn (Fig.~\ref{fig:moonletmigrate}). Aside from complicating the
  measurement of a moonlet's orbital elements within our simulations, 
  this motion may complicate interpretations of non-Keplerian motion 
  of moonlets within the ring \citep[e.g., Bl\'{e}riot,][]{tis10}.

\item {\bf The `young' Daphnis.} If some external physical mechanism 
  cannot maintain Daphnis in its current orbit, it must be a recent, 
  $\sim 10^3 - 10^4$~yr, addition to the ring.  Migration theory predicts 
  rapid damping in $(e, i)$ followed by fast migration. Instruments on
  board {\it Cassini} can measure the predicted damping in height above 
  the ring plane.
  
\item {\bf The `old' Daphnis.} If orbital resonances or stirring by distant
  moons can maintain Daphnis in its current orbit, it may be very old.  From 
  our simulations, we identify sources of stirring
  which might maintain Daphnis' orbit for timescales exceeding $10^4$~yr.
  Measuring oscillations in $(a, e, i)$ can place strong constraints on
  this mechanism.
  
\item {\bf Pan.}  With $\rhill =19$~km at an orbital distance of
  133,584~km, this small moon orbits within the 300~km-wide Encke
  gap~\citep{cuz85,jac08}.  Pan's Hill radius is very close to
  $\rfast$.  As with Daphnis, orbital resonances and stirring from
  distant moons probably prevent any form of migration. In addition,
  these effects may operate to help maintain the Encke gap, whose
  edges are relatively far, $\sim 8 \rhill$, from Pan. The torques 
  from ring material in this configuration  are too weak to cause 
  measurable radial drift.

\item {\bf Atlas.} Orbiting within the Roche Division at the outer edge of 
  the A ring, Atlas' Hill radius, $\rhill$ = 22~km, is also close to $\rfast$. 
  Although it is too far away from ring material to have any chance at 
  migrating, Atlas may have migrated out through the ring to its present 
  location. As with Pan and Daphnis, understanding the internal physics 
  of ring material and the interactions with distant moons might yield
  clues about its origin and recent history.

\item{\bf Pan and Atlas.} For both moons, detecting or limiting variations
  in $(a, e, i)$ can provide important constraints on the physical
  mechanisms within the ring and interactions between ring particles, 
  these small moons, and distant moons.

\end{itemize}

To conclude, our analytic estimates and numerical simulations
demonstrate that Saturn's rings are an important laboratory for
testing migration theory.  Current results are encouraging. Within the
A~ring, we can explain aspects of the radial and size distributions of
propellers and we identify plausible mechanisms for limiting migration
in Daphnis. We also identify several important observational tests
which can be accomplished with current satellites.

Aside from including additional constraints on ring properties as new
observations provide them, future studies of ring dynamics will eventually 
require including the ring self-gravity directly.  Self-gravity adds to 
the rich dynamical structure of the rings \citep[e.g.,][]{lew09}, which 
likely has interesting consequences for the migration of moons and moonlets 
(e.g., Fig.~\ref{fig:moonletmigrate}).  

\acknowledgements

We are grateful for advice and comments from M.~Geller. We thank A.~Youdin 
for suggesting important improvements to the manuscript.  Numerous helpful 
comments from the referee, M. Tiscareno, significantly improved the text.
Portions of this project were supported by {\it NASA's } {\it Astrophysics 
Theory Program} and the {\it Origin of Solar Systems Program} through grant 
NNX10AF35G and the {\it Outer Planets Program} through grant NNX11AM37G. We 
also acknowledge generous allotments of computer time on the NASA `discover' 
computer cluster operated by the National Center for Climate Simulation.

\appendix
\section*{Appendix}

The onset of fast migration relies on the presence of some ring material 
within $\sim 4\rhill$ of a small moon.  For a moon within a gap, fast 
migration proceeds only when the half-width of the gap is $\lesssim 4
\rhill$. Because the relationship between the moon and the size of the
gap it creates is critical to migration,  we consider several approaches 
to assess this relationship.


In the torque-balance approach \citep[e.g.,][]{lis81}, the viscous torque 
on ring material as it diffuses into a gap of width $\Delta a$ is set equal 
to the torque at the gap edge applied by the satellite. From weak-scattering 
theory \citep{lin79}, this balance yields a simple relationship between the 
gap half-width and the satellite's Hill radius:
\begin{equation}
\label{eq:torqbal1}
\Delta a \sim 2 \left(\frac{G M}{a^5 \nu^2}\right)^{1/6} \rhill^2 ~ .
\end{equation}
If the time to fill the gap is the viscous time (e.g., $\tfill$ in
eq.~[\ref{eq:tfill}] in the main text), it is possible to eliminate the 
viscosity in this expression and derive a minimum gap size, as in 
eq.~(\ref{eq:rgap}) for $\rgap$.

An alternative approach equates the radial distance ring material can diffuse 
in a synodic period, $\sqrt{\nu\tsynodic}$, with the displacement in semimajor 
axis a particle receives after every close encounter with the satellite 
(proportional to the right-hand-side of eq. (\ref{eq:torqbal1}) above; see also 
eq.~[5] of Bromley \& Kenyon 2011b). This balanced is achieved at an orbital 
distance from the satellite 
\begin{equation}
\label{eq:torqbal2}
\Delta a \sim \left(\frac{G M }{a^5 \nu^2}\right)^{1/18} \rhill^{4/3} ~ .
\end{equation}
This result is similar to the result in eq.~(\ref{eq:torqbal1}).

More detailed treatments derive some form of the radial diffusion equation 
across the gap and solve explicitly for the surface density distribution and 
the half-width of the gap \citep[e.g.,][]{hour1984,raf01}. Although these 
solutions yield much better estimates for the surface density distribution,
results for the half-width of the gap differ from the simpler treatments by 
less than a factor of two. Given the uncertainties, either eq.~(\ref{eq:torqbal1})
or eq.~(\ref{eq:torqbal2}) provides a reasonable approximation to the half-width.

To apply these two results to Saturn's A ring, we adopt the nominal parameters 
in eqs.~(\ref{eq:Sigmafid})--(\ref{eq:nuradfid}) from the main text and derive
\begin{equation}
\label{eq:torqbal3}
\Delta a \sim 0.97 \left[\frac{\rhill}{\rkm}\right]^2
\left[\frac{\noo85}{\nurad}\right]^{1/3}
\left[\frac{\a130}{a}\right]^{5/6} \ {\rm km}
\end{equation}
for balancing the torques and
\begin{equation}
\label{eq:torqbal4}
\Delta a \sim 1.4 \left[\frac{\rhill}{\rkm}\right]^{4/3}
\left[\frac{\noo85}{\nurad}\right]^{1/9}
\left[\frac{\a130}{a}\right]^{5/18} \ {\rm km}
\end{equation}
for balancing scattering and viscous diffusion.  For small moons capable of 
fast migration, $\rhill \approx$ 2--5~km, these approaches yield similar 
results for $\Delta a$. For much larger moons, torque balance consistently
yields a larger half-width.

For the rest of this section, we focus on half-widths derived by balancing
scattering and viscous diffusion.  Requiring $\Delta a \gtrsim 4 \rhill$ 
yields the maximum Hill radius for a satellite capable of drawing material 
from the gap edges, $\sim$ 23~km, when scattering balances viscous diffusion.  
This limit is comparable to $\rfast$. 

In principle, moons with $\rhill \gtrsim$ 23~km can prevent material from 
encroaching within $4\rhill$. These moons migrate slowly in the type II mode.
Because gravitational scattering falls off so steeply with orbital distance 
from the satellite, viscous diffusion constantly tries to drive material 
into the gap. If diffusion produces a higher surface density and viscosity 
at the edges of a gap, the gap will shrink, enabling the satellite to draw 
material away from the gap, possibly initiating fast migration. 

To compare the simple theory with Daphnis in the Keeler Gap and Pan in
the Encke Gap, we adopt $\rhill \approx$ 5~km for Daphnis and $\rhill
\sim 20$~km for Pan. For Daphnis, the balance between scattering and
viscous diffusion occurs at $\Delta a \sim 12$~km, somewhat smaller
than the observed half width, $\sim$ 13--20~km, of the Keeler gap
\citep{wei09}.  For Pan, the balance is at $\Delta a \sim$ 75~km, much
smaller than the observed half-width, $\sim$ 160~km, of the Encke
Gap. Based on this analysis, we identify Daphnis as a good candidate
for fast migration. Pan is a poor candidate for fast migration, at
least under current conditions in the Encke Gap. However, barring
other effects such as resonances with distant moons, Pan may be unable
to halt long-term diffusion of ring material to within $4\rhill$.

In our calculations, large moons achieve a balance between viscous diffusion 
and gravitational scattering. The numerical results support the limits 
derived in eq.~(\ref{eq:torqbal4}).  Moons with $\rhill \gtrsim$ 20--25~km 
can migrate in the type II mode. Smaller moons with $\rhill \approx$ 2--20~km 
may undergo fast migration.

\newpage


\begin{figure}[htb]
\centerline{\includegraphics[width=5.5in]{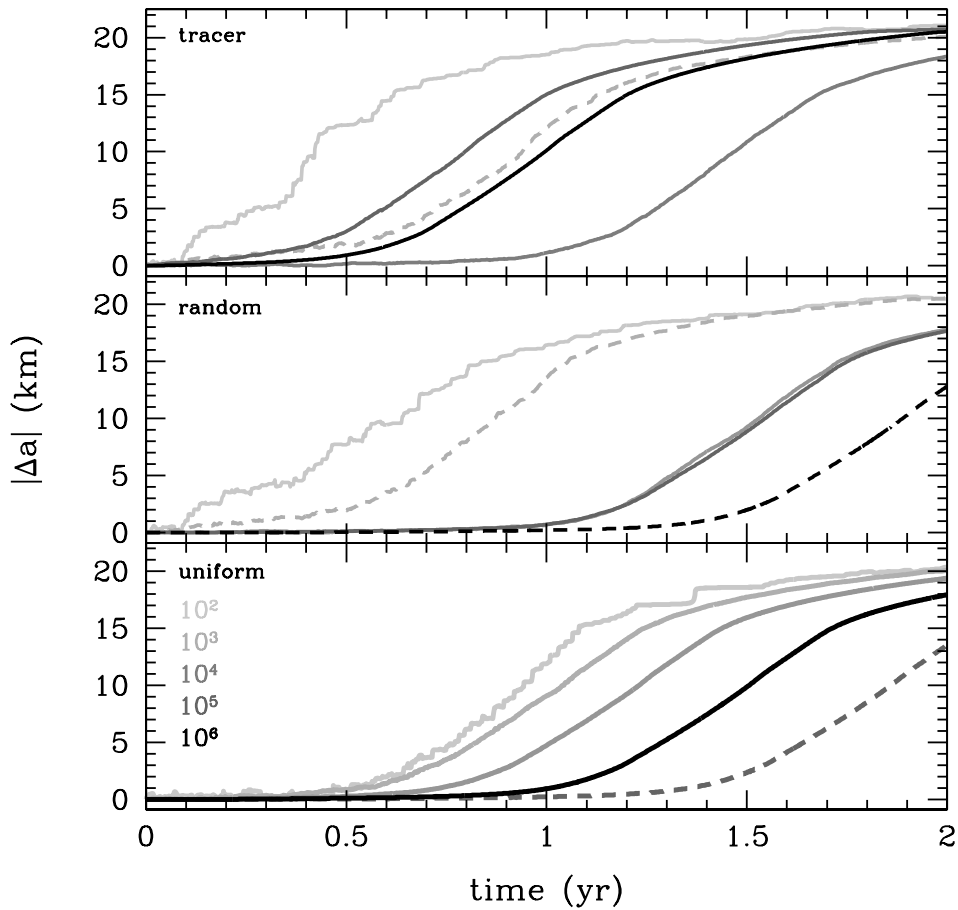}}
\caption{Fast migration of small moons in a ring of massive particles
  with $\Sigma = 40$~g/cm$^2$, a total annular width of 56~km, and a
  gap with a halfwidth of 14~km, which covers the corotation zone.
  Simulations begin with the moon precisely centered in the
  gap. Each curve represents the change in orbital distance, $\Delta
  a$, of a moon with $\rhill = 7$~km and an initial semimajor axis
  of $130,000$~km. In each simulation, the particles in the ring have
  identical mass, set to $10^{-2}$ (lightest shading), $10^{-3}$,
  $10^{-4}$, $10^{-5}$, or $10^{-6}$ (darkest shading) times the mass
  of the moon.  Solid (dashed) lines indicate inward (outward)
  migration. Although migration is more stochastic with larger ring
  particles, the migration rate is nearly independent of the masses of
  ring particles.  When a moon reaches the edge of the ring,
  migration ceases. 
\label{fig:ringxx}}
\end{figure}

\begin{figure}[htb]
\centerline{\includegraphics[width=5.5in]{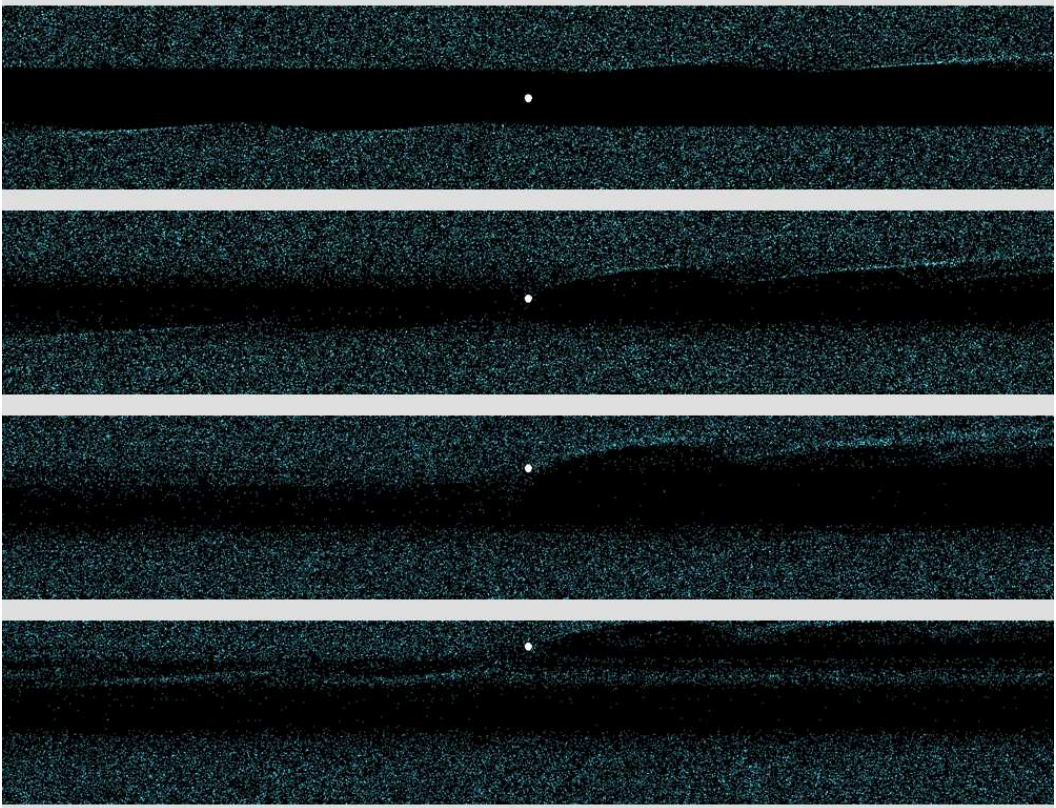}}
\caption{Snapshots of small moon migration in a closing gap.  The top
  image showa the location of a Daphnis-size moon ($\rhill = 4.9$~km,
  the central white disk of 2.3~km in radius, the approximate physical
  size of Daphnis) at Daphnis's orbital distance (136,505~km) when it
  is initially on a circular orbit in a pristine ring with uniform
  surface density $\Sigma = 40$~g/cm$^2$, centered on a clear annulus
  the size of the Keeler gap (full width of 40~km).  The ring
  particles (masses equal to 0.1\% times that of the
  satellite) are indicated with a brightness that correlates with
  density. The lower three images show views from the same orbital
  distance from Saturn but at times equal to 46.6~yr (second from
  top), 48.2~yr, and 49~yr. Radial diffusion ($\nurad = 85$~cm$^2$/s)
  causes `the ring to slowly creep into the gap, but once particles
  get close, fast migration initiates and the moon heads toward
  Saturn.  Each image represents an area of 800~km in width and 120~km
  in height, oriented so that Saturn is located toward the top of the
  page.  The ``exposure time'' for each image is several orbital
  periods, to enhance the structures in the ring.
\label{fig:daphnis2x96}}
\end{figure}

\begin{figure}[htb]
\centerline{\includegraphics[width=5.5in]{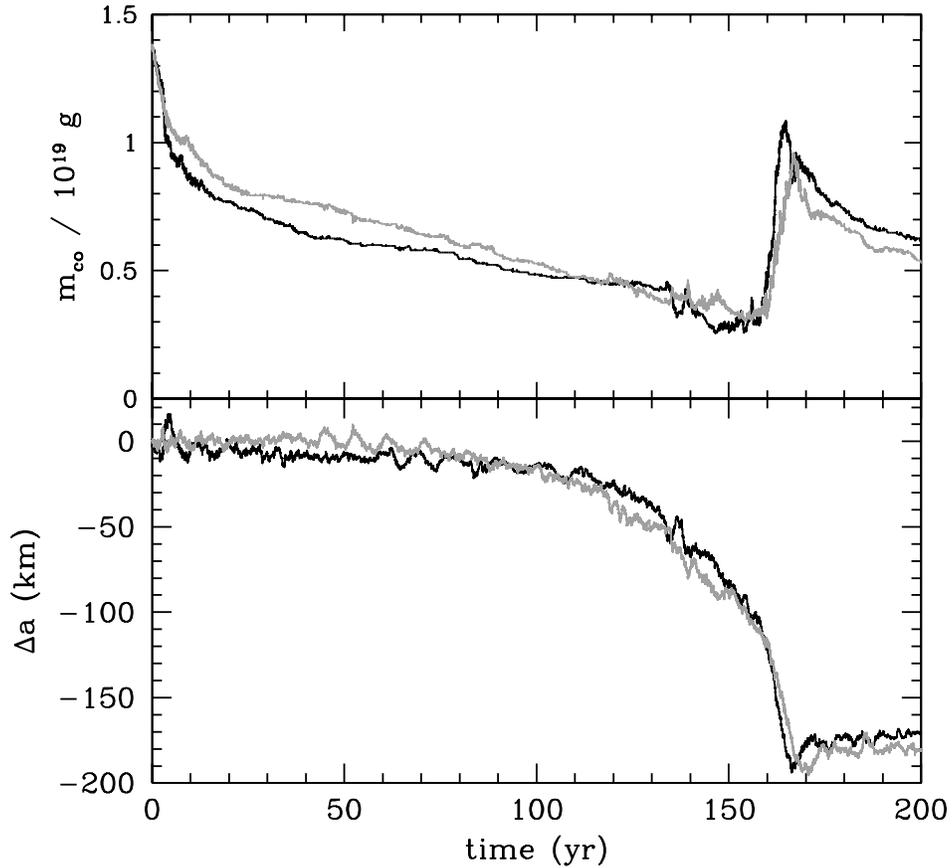}}
\caption{The disk mass in the corotation zone and the radial drift of
  a small moon with $\rhill = 14$~km.  The starting orbital distance is
  130,000~km. The upper plot shows the instantaneous disk mass $m_{\rm
  co}$ inside an annulus of half-width $2\rhill$; the lower plot is
  the orbital distance of the moon relative to its starting position. 
  The black curve refers to calculations for the moon and ring alone; 
  the light curve indicates results for calculations which include 
  Saturn's prominent moons (Fig.~\ref{fig:fullring}).  
  In both instances, migration stops after $\sim 170$ years, when the 
  moon hits the inner edge of the simulated disk.
\label{fig:mhorse}}
\end{figure}

\begin{figure}[htb]
\centerline{\includegraphics[width=5.5in]{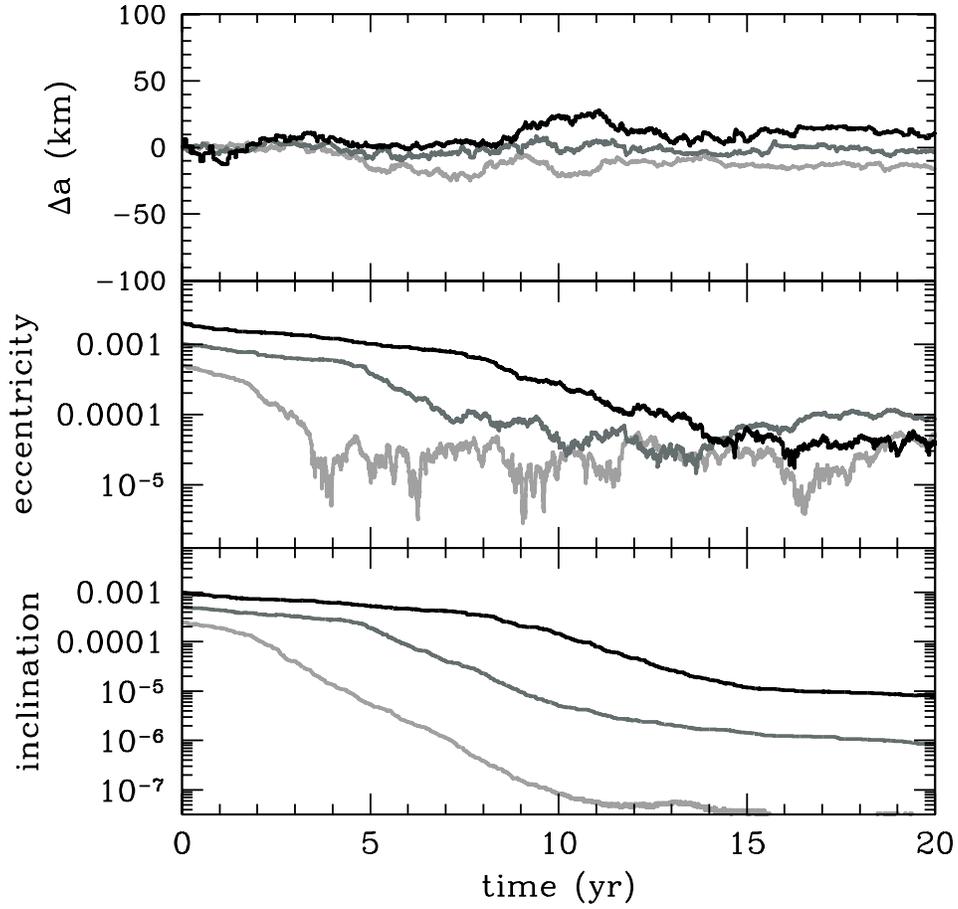}}
\caption{Orbital settling of captured small moons. Simulations of
moons with Hill radii of 10~km (light gray curves), 
14~km (dark gray curves), and 18~km (black curves) show the 
evolution of orbital elements in an initially uniform disk 
with A ring parameters. In all cases, the time for each moon 
to become embedded in the disk is very short.
\label{fig:capture}}
\end{figure}

\begin{figure}[htb]
\centerline{\includegraphics[width=5.5in]{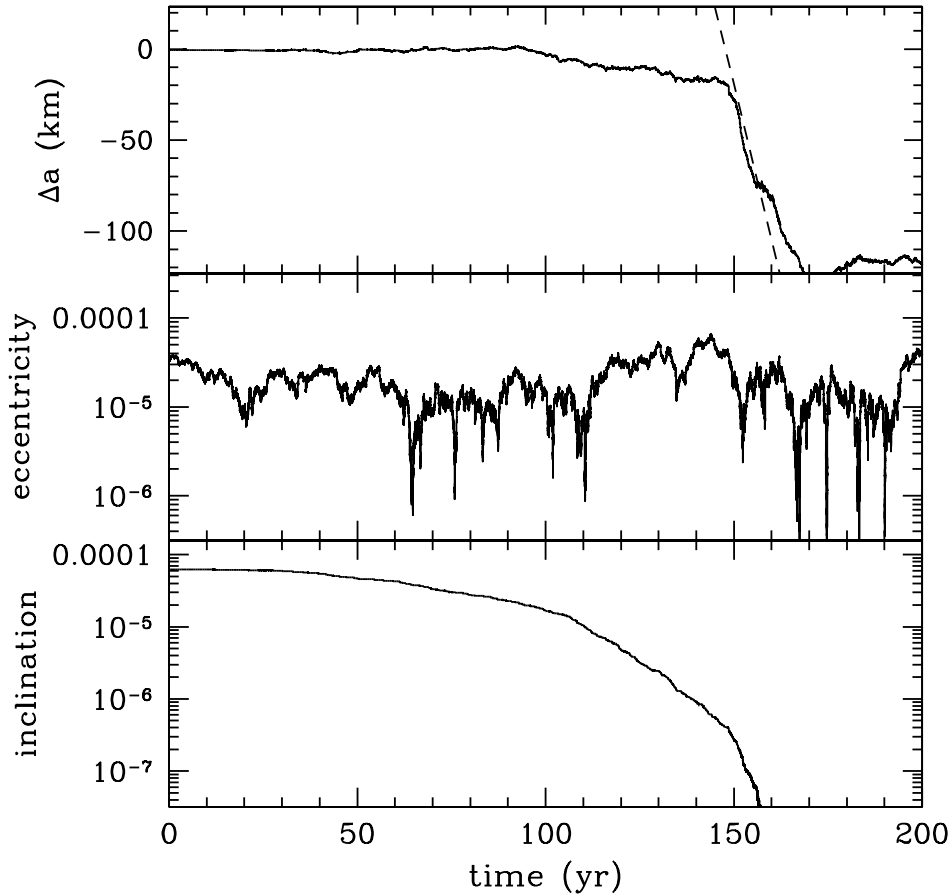}}
\caption{The onset of fast migration of a captured moon.
The traces show the change in semimajor axis (upper plot), eccentricity
(middle plot) and inclination (lower plot) of a small moon with $\rhill = 6.5$~km
in a disk with $\Sigma = 20$~g/cm$^3$.
The initial orbit of the moon is the same as for Daphnis;
the disk properties are modified for computational feasibility.
The disk has an 80~km-wide gap (similar to the Keeler gap along Daphnis'
orbit in terms of the moon's Hill radii). The disk particles'
eccentricity and inclination damp as expected for the A ring. The 
viscosity, however, is turned up by a factor of 15, thus reducing the
viscous spreading time in the gap by a factor of 3 compared to the
Keeler gap. The simulation shows that the eccentricity damps quickly,
but migration does not begin until the inclination is also small. After
migration starts, its rate is comparable to the theoretical prediction
(dashed curve), and it halts once the moon nears the edge of
the simulated disk.
\label{fig:daphnist100}}
\end{figure}

\begin{figure}[htb]
\centerline{\includegraphics[width=5.5in]{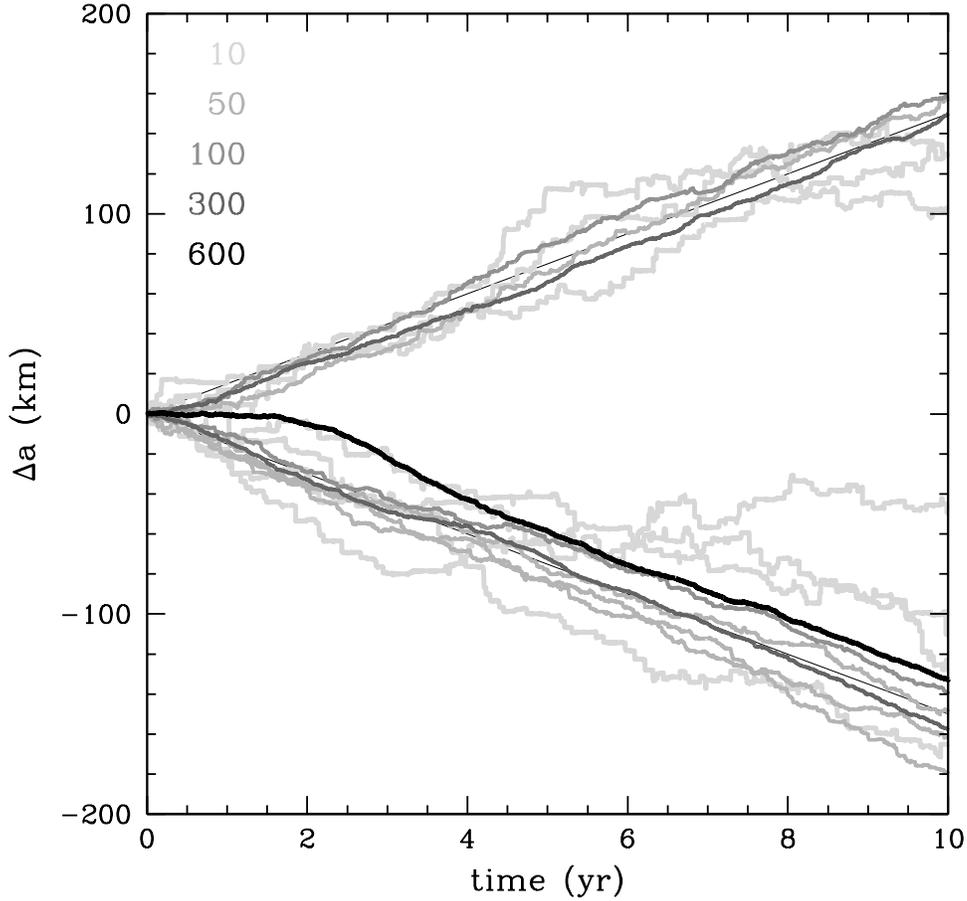}}
\caption{Fast migration of small moons in a ring of massive particles with
  $\Sigma = 30$~g/cm$^2$. Each curve represents the change in orbital
  distance, $\Delta a$, of a moon with $\rhill = 14$~km that
  initially has a semimajor axis of $130,000$~km.  In each simulation
  the particles in the ring have identical mass, and are set to either
  10, 50, 100, 300 and 600 times smaller than the mass of the moon.
  The darkest curve shows the case with the most extreme mass raio
  (600:1); lighter shades have progressively less extreme mass
  ratios. The lightest curves, corresponding to runs where the
  particle masses are 10\% of the moon mass, show stochastic
  behavior.
\label{fig:massrat}}
\end{figure}

\begin{figure}[htb]
\centerline{\includegraphics[width=5.5in]{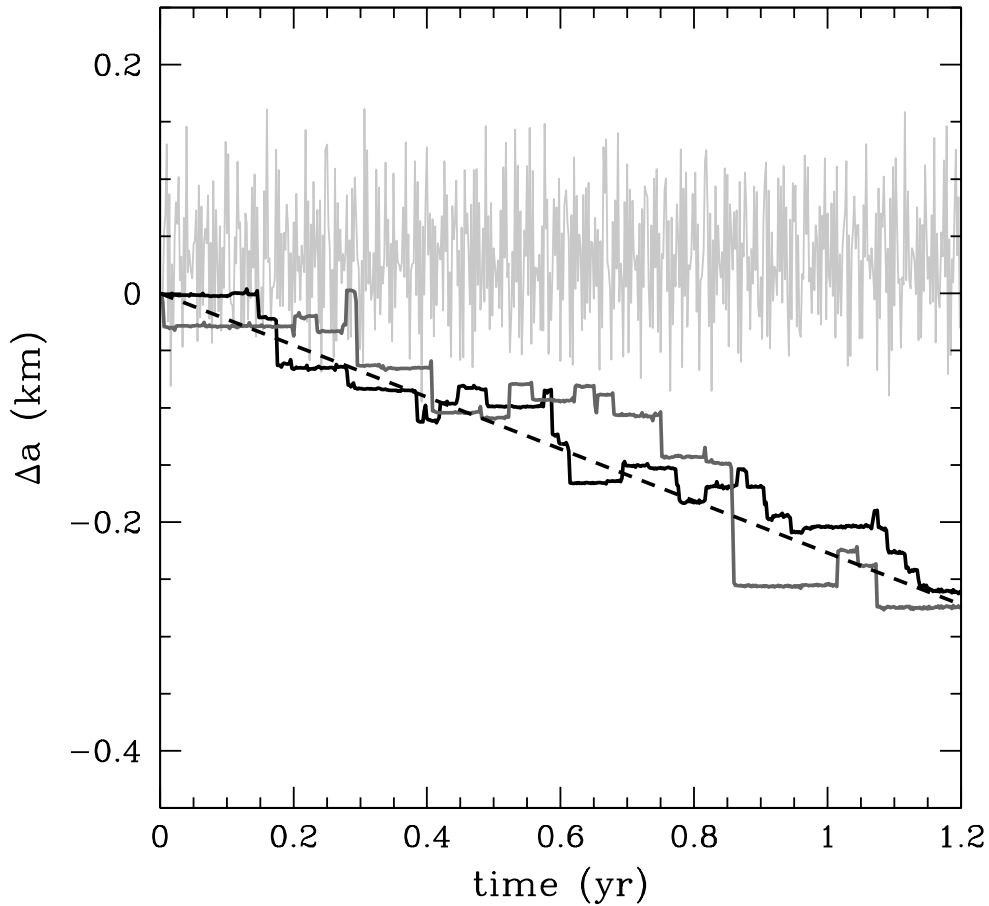}}
\caption{Fast migration of a small moonlet in the presence of Saturn's
  massive moons. In these simulations a moonlet with $\rhill =
  0.7$~km starts off at an orbital distance of 130,000~km in a ring
  with surface density $\Sigma = 0.1$~g/cm$^3$.  The black curve
  corresponds to ring particles that are 1\% of the moonlet mass,
  while the dark gray curve shows results for ring particles that are
  twice as massive.  The corotation zone is initially clear, and no
  viscous diffusion can fill in the gap. Thus, fast migration ensues,
  and there is a subsequent change in orbital distance, $\Delta a$
  (solid black curve; the dashed line shows the theoretical
  prediction). Note that the true orbital distance (not shown) is
  given by the sum of $\Delta a$ and the orbital fluctuations that
  result from orbital ``jostling'' of the moonlet by the distant
  moons.  We estimate this jostling using a moonlet with identical
  starting conditions as the other simulations, but without any ring
  particles (light gray curve).
\label{fig:moonletmigrate}}
\end{figure}

\begin{figure}
\centerline{\includegraphics[width=5.5in]{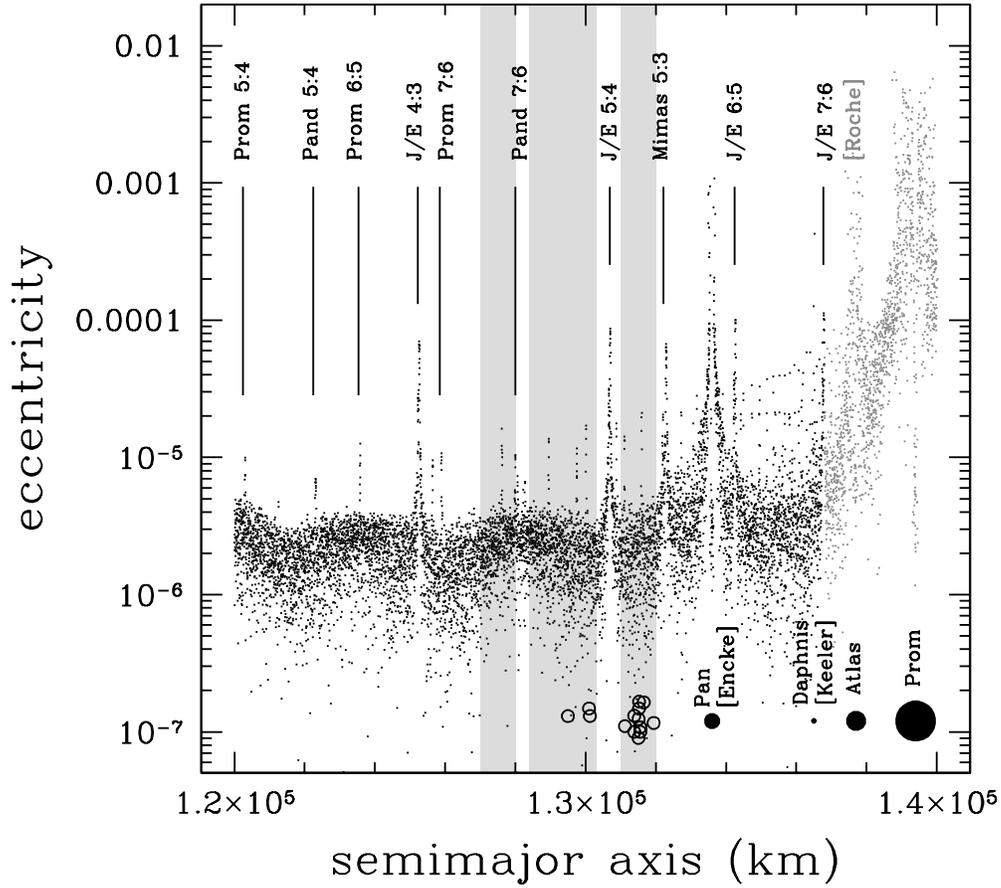}}
\caption{The eccentricity of simulated ring particles as a function of
  semimajor axis in the presence of massive moons. The simulation
  started from cold initial conditions and evolved tracer particles
  over a 4-year period. Some of the strongest resonances are labeled;
  off-resonance pumping of eccentricity is also apparent throughout
  the ring. The black circles designate moons; the size
  of each circle is suggestive of the relative mass. The open circles
  are a select sample of propeller moonlets from \citet{tis08}.
  None of these symbols are intended to quantify eccentricity, just the
  semimajor axis. The gray bands are the zones where propeller moonlets
  seem to reside exclusively. The figure also labels the location of the
  Encke and Keeler gaps, as well as the inner edge of the Roche Division
  (containing the light gray points). 
\label{fig:fullring}}
\end{figure}

\begin{figure}[htb]
\centerline{\includegraphics[width=5.5in]{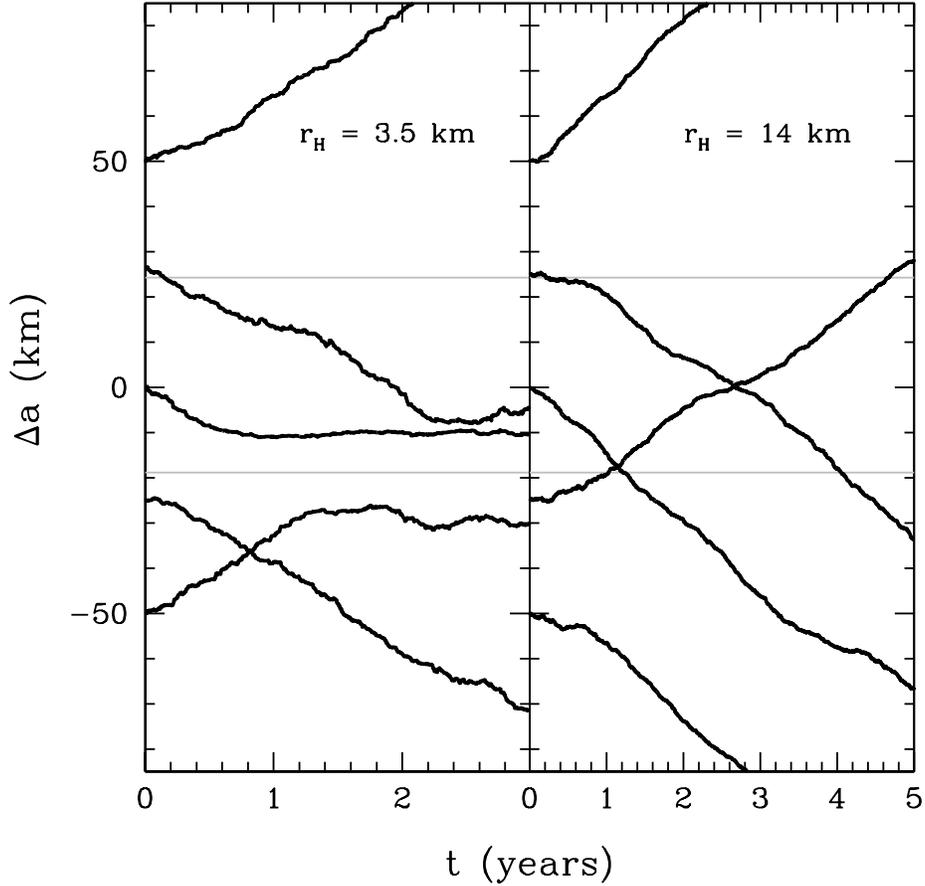}}
\caption{Fast migration in the presence of resonances from distant
  moons Janus and Epimetheus.  The two panels show the output of
  simulations with isolated small moons on circular orbits at a position
  near the resonances. The black-colored traces show the radial
  position of each moon $\Delta a = 0$~km relative to an orbital
  distance of $a = 130,500$~km. For reference, the light grey lines
  give the location of the resonances from Janus and Epimetheus (upper
  and lower lines, respectively).  The five runs in the left panel
  show fast migration of a smaller ($\rhill = 3.5$~km) moon; the
  runs in the right panel are for a larger moon ($\rhill =
  14$~km). For each set of runs the initial conditions are identical,
  except that the moon and ring are radially offset by a differing
  amount. The results indicate that the smaller moons cannot cross
  the resonance; the larger ones can pass through.
\label{fig:avtjanepi}}
\end{figure}

\begin{figure}[htb]
\centerline{\includegraphics[width=5.5in]{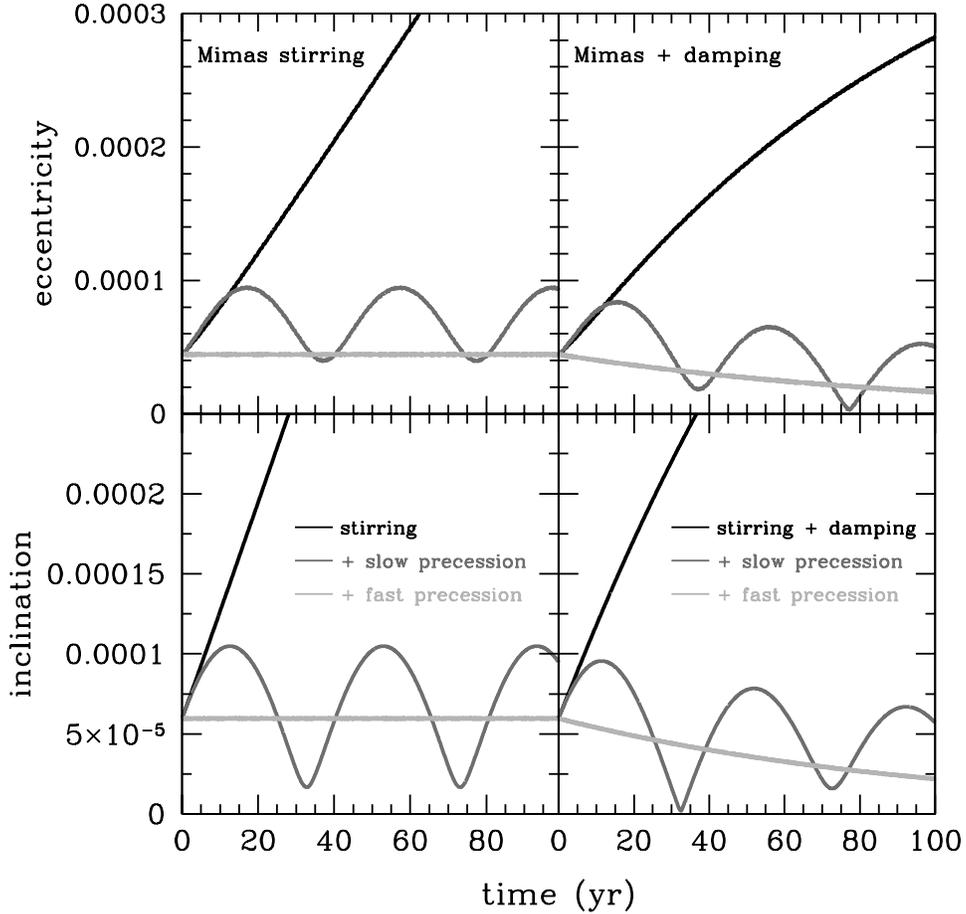}}
\caption{Stirring of a single massless ring particle by a distant moon 
  (Mimas). The ring particle has initial orbital elements similar to 
  Daphnis.  Left panels: stirring by Mimas without collisional damping.
  Right panels: stirring by Mimas with collisional damping. In each panel,
  the black curves show the growth in $e$ and $i$ when Saturn and Mimas 
  have a spherically symmetric gravitational potential. Stirring grows
  continuously. When Saturn's oblateness is included in the calculations,
  the ring particle precesses rapidly. Stirring and precession yield a
  rough equilibrium in $e$ and $i$ as expected for a ring particle. 
  When precession is slowed by a factor 
  of 100, the variations in $e$ and $i$ become periodic due to the precession
  of the ring particle's angular momentum vector precesses (dark gray curves). 
\label{fig:stir}}
\end{figure}

\begin{figure}[htb]
\centerline{\includegraphics[width=5.5in]{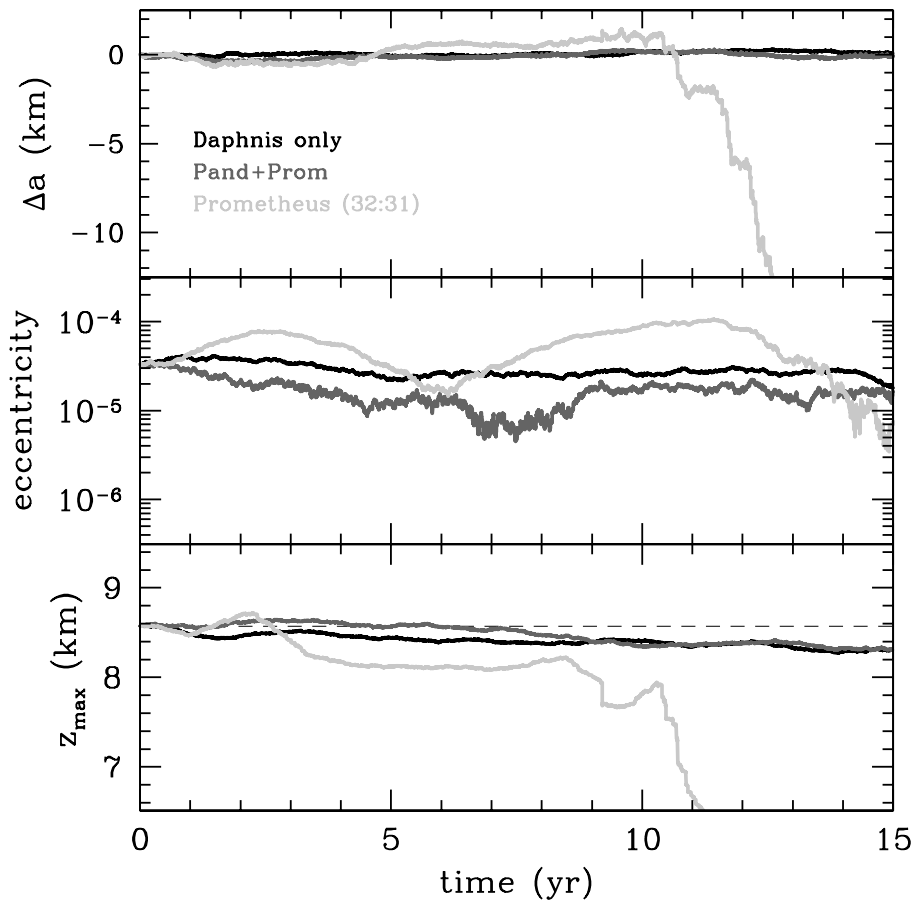}}
\caption{Simulations of Daphnis' orbit: damping, stirring and
  resonances.  The traces show changes in the semimajor axis (upper
  panel), eccentricity (middle panel) and maximum altitude above the
  ring plane (lower panel, the inclination is $i \approx z_{\rm max} /
  a$) of a simulated moon with $\rhill = 2.6$~km at $a =
  136,505$~km. The initial eccentricity and inclination are similar to
  those of Daphnis \citep{jac08}.  To help isolate physical effects,
  the oblateness of Saturn is not included in these simulations (see
  Fig.~\ref{fig:stir}).  The curves illustrate the evolution of (i) a
  small moon embedded in the Keeler gap (black curve), (ii) a small moon
  with its position adjusted to coincide with the 32:31 resonance of
  Prometheus (light gray curve), and (iii) a small moon, off-resonance but
  stirred by Prometheus and Pandora (dark gray curve).  Without
  resonance or interactions with distant moons, Daphnis settles into
  the ring ($\sim 10^2 - 10^3$ yr) and then begins fast migration. In
  a single resonance, Daphnis is scattered out of the Keeler gap and
  begins to migrate. Evolution within the Pandora 18:17 resonance is
  similar but on a longer timescale. Off resonances, the orbit is
  stable on timescales of at least $10^4$~yr before Daphnis begins to
  settle into the ring.
\label{fig:daphnis}}
\end{figure}

\begin{figure}[htb]
\centerline{\includegraphics[width=5.5in]{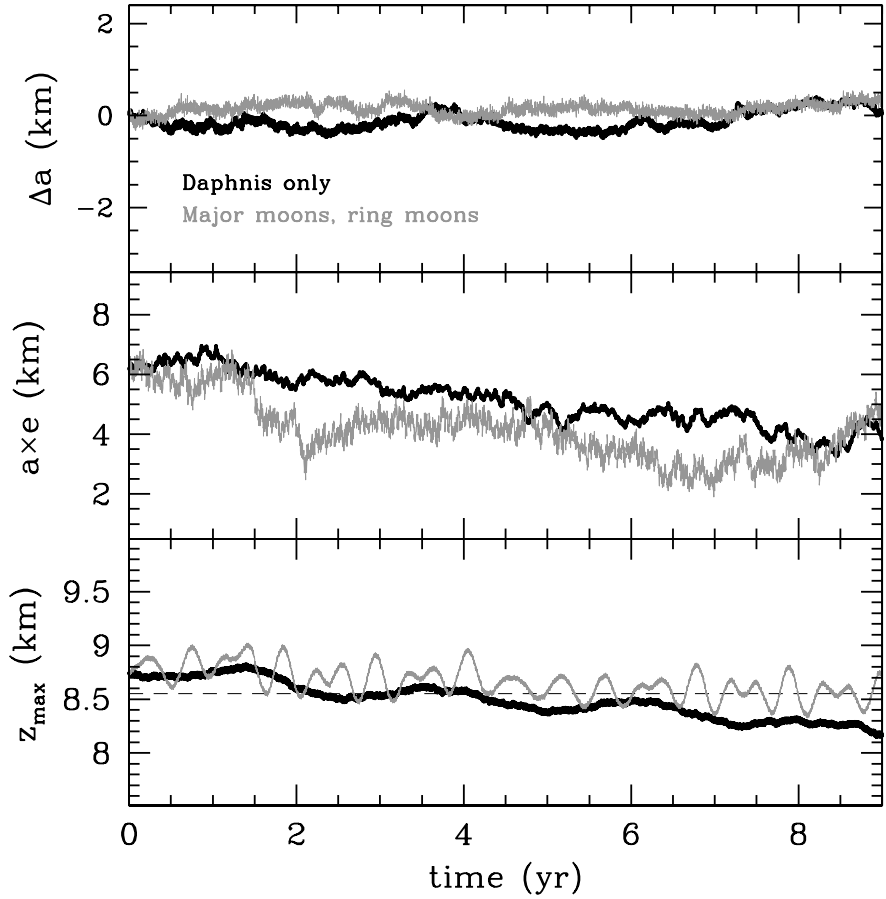}}
\caption{Simulated evolution of Daphnis' orbit when stirred by Saturns
  moons.  The moon orbits within a gap in surface density with a
  density contrast and width similar to the Keeler gap.  As in 
  Figure~\ref{fig:daphnis}, the black curve shows changes in the
  moon's orbital elements solely from interaction with the A-ring;
  over the course of 6000 orbital periods, Daphnis slowly settles in
  the ring and eventually begins to migrate rapidly. The light curve
  shows the evolution when Daphnis interacts with ring particles and
  all of the moons of Saturn. Using initial phase-space positions from
  JPL's Horizons calculator \citep{gio96}, Saturn's moons produce
  quasi-periodic fluctuations in the inclination, which prevent
  Daphnis from settling into the ring plane. Thus, interactions with
  distant moons plausibly prevent fast migration of Daphnis through
  the A~ring.
\label{fig:daphnisall}}
\end{figure}

\end{document}